\documentclass[prb,aps,epsf,twocolumn,showpacs]{revtex4-1}
\usepackage[pdftex]{graphicx}
\usepackage{dcolumn}
\usepackage{bm}
\usepackage{epsfig}
\usepackage{latexsym}
\usepackage{amsmath}
\usepackage{amssymb}
\usepackage{color}
\usepackage{array}
\usepackage{float,subfigure}

\renewcommand{\(}{\left(}
\renewcommand{\)}{\right)}
\renewcommand{\[}{\left[}
\renewcommand{\]}{\right]}
\renewcommand{\v}[1]{\mathbf{#1}} 

\newcommand{\bs}{\boldsymbol}
\newcommand{\p}{\parallel}
\newcommand{\pp}{k_{\parallel}} 
\newcommand{\ppp}{k_{\perp}}

\newcommand{\kf}{k_{F}}
\newcommand{\ko}{k_{0}}
\newcommand{\be}{\begin{equation}}
\newcommand{\bsplit}{\begin{split}}
\newcommand{\ee}{\end{equation}}
\newcommand{\zb}{z_{b}}
\newcommand{\mone}{m_{1}}
\newcommand{\mtwo}{m_{2}}
\newcommand{\mo}{m_{0}}
\newcommand{\psid}{\psi^{\dag}}
\newcommand{\eps}{\epsilon}

\newcommand{\lam}{\lambda}

\newcommand{\zhat}{\hat{z}}

\newcommand{\kpp}{k_{\perp}}
\newcommand{\kp}{k_{||}}

\newcommand{\OM}{\Omega}
\newcommand{\om}{\omega}
\newcommand{\QPP}{q_{\perp}} 
\newcommand{\QP}{q_{||}}

\newcommand{\vj}{v_{j}}
\newcommand{\mj}{m_{j}}
\newcommand{\vjp}{v_{j'}}
\newcommand{\mjp}{m_{j'}}
\newcommand{\Twopic}{(2\pi)^{3}}

\newcommand{\noin}{\noindent}
\newcommand{\etap}{\eta'}

\newcommand{\EM}{\hat{e}_{M}}
\newcommand{\vd}{v_{D}}

\begin{document}
\title{Stable non-Fermi liquid phase of itinerant spin-orbit coupled ferromagnets}
\author{Yasaman Bahri and Andrew C. Potter}
\affiliation{Department of Physics, University of California, Berkeley, CA 94720, USA}
\date{\today}
\begin{abstract}

Direct coupling between gapless bosons and a Fermi surface results in the destruction of Landau quasiparticles and a breakdown of Fermi liquid theory. Such a non-Fermi liquid phase arises in spin-orbit coupled ferromagnets with spontaneously broken continuous symmetries due to strong coupling between rotational Goldstone modes and itinerant electrons.  These systems provide an experimentally accessible context for studying non-Fermi liquid physics. Possible examples include low-density Rashba coupled electron gases, which have a natural tendency towards spontaneous ferromagnetism, or topological insulator surface states with proximity-induced ferromagnetism.
Crucially, unlike the related case of a spontaneous nematic distortion of the Fermi surface, for which the non-Fermi liquid regime is expected to be masked by a superconducting dome, we show that the non-Fermi liquid phase in spin-orbit coupled ferromagnets is stable. 

\end{abstract}

\maketitle

The vast majority of gapless quantum phases are Fermi liquid metals with asymptotically sharp electronic quasiparticles. A notable class of exceptions occurs when a finite density of fermions couples directly to a gapless bosonic mode. Such a situation is expected to arise in exotic contexts such as quantum critical points in metals\cite{MaxIsingI, MaxIsingII} and quantum spin liquids where emergent gauge fields couple to a Fermi surface of fractional spinon or composite fermion excitations.\cite{IoffeLarkin, HLR, LeeLee, SenthilMott} In this paper, we explore a more conventional context for non-Fermi liquid (NFL) physics: metals with spontaneously broken rotational symmetry.\cite{Oganesyan}

Spontaneous breaking of a continuous symmetry results in gapless bosonic modes, a necessary ingredient for NFL physics.  However, Goldstone modes ordinarily decouple from other excitations at low energies (see Ref.~\citenum{Haruki} for a general criterion). In particular, rotational Goldstone modes \emph{do} couple strongly to the Fermi surface so long as there is no concomitant translational symmetry breaking.\cite{Haruki}
In this case, the rotational modes strongly excite particle-hole pairs, become overdamped, and (in two dimensions) destroy the coherence of electronic quasiparticles. This phenomena was originally discussed in the context of spontaneous nematic distortions of the Fermi surface.\cite{Oganesyan,Congjun1} However, it was recently shown\cite{MaxPairing} that this system has a strong tendency towards superconductivity that preempts the onset of NFL behavior and obscures the potential NFL phase.  This raises the general question: can stable NFL behavior result from an overdamped rotational mode?  


In this paper, we provide a simple class of examples for which superconductivity does not preempt NFL behavior: itinerant ferromagnets (FM) with broken continuous rotation symmetry and spin-orbit coupling (SOC). \cite{CenkeTIFM} A central result of our paper is that for SOC FMs, the breaking of both time-reversal and inversion symmetries disrupts Cooper pairing and renders the NFL phase stable to superconductivity (and indeed to other subsidiary symmetry breaking orders that could disrupt the NFL phase).

\begin{figure}[ttt]
\begin{center}
\includegraphics[width = 3.5in]{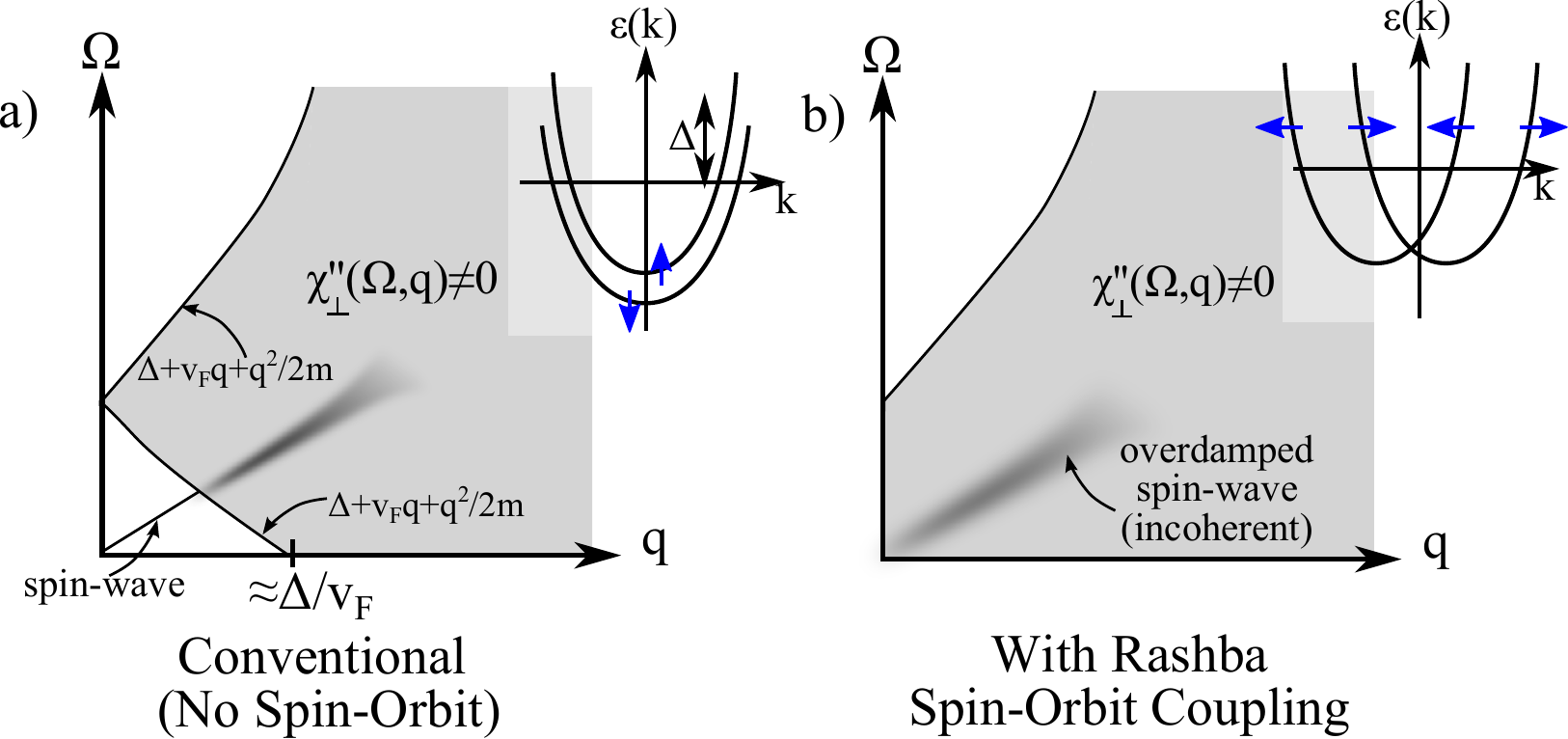}
\end{center}
\vspace{-.2in}
\caption{(a) The imaginary part of the transverse spin-susceptibility $\chi''$ of a conventional ferromagnet (FM) (dispersion shown as an inset) has a gap at small frequency, $\Omega$, and wave vector, $q$, in which a sharp spin wave excitation can exist.  In contrast, the Rashba coupled FM (b) has finite $\chi''$ for small $(\Omega,q)$ leaving only an overdamped spin wave that is strongly coupled to the electrons, resulting in non-Fermi liquid (NFL) behavior. 
}
\vspace{-.2in}
\label{fig:TransverseSusceptibility}
\end{figure}

The origin of NFL physics in SOC FMs was first pointed out in Ref. \citenum{CenkeTIFM} and can be understood by comparing to ordinary FMs without SOC.  There, FM order splits the initially spin degenerate Fermi surface into two spin-polarized Fermi surfaces (FSs) (Fig. \ref{fig:TransverseSusceptibility}). Consequently, fluctuations of the order parameter create electronic spin-flip excitations only by supplying either nonzero energy or momentum. Hence, there are no electron-hole excitations at small energies and wave vectors, and in this region FM spin waves effectively decouple from the electronic particle-hole continuum and exist as sharp, undamped excitations. This dynamical decoupling of low-energy order parameter fluctuations is typical of Goldstone modes of a broken continuous symmetry.\cite{Haruki} 

By contrast, with SOC the electron spins are tied to momentum and are not fully polarized by the FM order so that spin-flip excitations can be made at arbitrarily low energy and momenta over nearly the entire Fermi surface (Fig.~\ref{fig:TransverseSusceptibility}). Hence, spin wave excitations cannot propagate without exciting electron-hole pairs, causing the spin wave modes to become strongly overdamped. In turn, electronic quasiparticles near the Fermi surface are strongly scattered by the overdamped spin waves, which now have softer dynamics scaling like $\Omega\sim q^3$.  This strong scattering destroys the integrity of Landau quasiparticles, leading to a non-Fermi liquid.

The SOC itinerant FM represents perhaps the simplest and most accessible context for exploring NFL behavior. Candidate materials include 1) 2D electron liquids with Rashba spin-orbit coupling and 2) topological insulator surface states doped with magnetic moments or coupled to ferromagnetic insulators.\cite{CenkeTIFM,FMTI} The former is particularly appealing because it has a natural tendency to Stoner-type FM at low densities due to a diverging density of states.\cite{Berg,RuhmanBerg}

The paper is organized as follows.  We begin by explaining the origin and character of non-Fermi liquid physics in SOC itinerant FMs.  Next, we discuss important considerations regarding crystal anisotropies and disorder, identify candidate materials, and describe several experimentally testable signatures of the NFL phase.  Finally, we analyze the stability of these NFL phases to the onset of potential superconducting and spin density wave (SDW) orders that would disrupt the NFL behavior.  Computations are done in the framework of the $(N,\epsilon)$ expansion of Ref.~\onlinecite{NEpsilon}, and details specific to the Rashba coupled systems may be found in the accompanying appendices.

\subsection{Relation to previous work}
The existence and basic physics of NFL phases in SOC FMs (in particular the TI surface state) was previously pointed out in Ref.~\onlinecite{CenkeTIFM}. This work expands upon that work by 1) analyzing the crucial issue of stability to subsidiary order, which likely obscures previously proposed NFL phases,\cite{MaxPairing} and 2) providing a more extensive discussion of experimental phenomenology and candidate materials. We also have a stronger focus on Rashba coupled systems, touched upon only briefly in Ref.~\onlinecite{CenkeTIFM}.


\section{Non-Fermi Liquid Phase}
Consider a system of spin-orbit coupled electrons described by (imaginary time) Lagrangian density $\mathcal{L} = \psid\[\partial_\tau+\mathcal{H}\]\psi$ that is invariant under combined spatial and spin rotations about $\bs{\hat{z}}$.  In this paper, we will focus on 
two dimensional electron liquids with Rashba SOC
\begin{align} \mathcal{H}_\text{R} = \frac{k^2}{2\mo}-\mu+ \alpha_R\bs{\hat{z}}\cdot\(\v{k} \times \bs{\sigma} \)
\end{align}
Here, we are chiefly interested in the low-density regime where the nonmagnetic Fermi surface has annular topology and an enhanced density of states naturally favors spontaneous ferromagnetism.\cite{Berg,RuhmanBerg} For completeness, however, we also present results for the high-density regime with a simply connected Fermi sea consisting of two concentric sheets, as well as the closely related toplogical insulator (TI) surface state with Hamiltonian:
\begin{align} \mathcal{H}_\text{TI} = v_D\bs{\hat{z}}\cdot\(\v{k} \times \bs{\sigma} \)-\mu
\end{align}

Suppose the system has spontaneous magnetization $\v{M}$ with an easy-plane (XY) anisotropy, whose long wavelength dynamics in the absence of coupling to electrons is governed by:
\begin{align} \mathcal{L}_M^{(0)} = |\partial_\tau \v{M}|^2+c^2|\nabla\v{M}|^2 \end{align}
and which couples to the electrons through a term
\begin{align} H_{e-M} = - \lambda_0 \v{M}\cdot \psi^\dagger\bs{\sigma}\psi 
\label{eq:Coupling}
\end{align} 
The XY-magnetization could represent either polarization of local moments, for instance due to magnetic doping or proximity to a ferromagnetic insulator, or the spin-polarization of the electrons themselves.\cite{Berg,RuhmanBerg}


Well below the magnetic ordering temperature $T_M$, only fluctuations in the direction, not the amplitude, of $\v{M}$ are important.  These can be parameterized by a Goldstone ``phase" field $\phi(\v{r},t)$ as follows: $\v{M}(\v{r},t) = M_0\[ \bs{\EM} \cos\phi+(\bs{\hat{z}} \times\bs{\EM})\sin\phi\] \approx M_0 \[ \bs{\EM} +(\bs{\hat{z}} \times\bs{\EM})\phi\]$, where we have linearized the fluctuations about the ordering direction.\cite{DiamagneticTerm}

\begin{center}
\begin{figure*}[t]
(a)\includegraphics[width=2in]{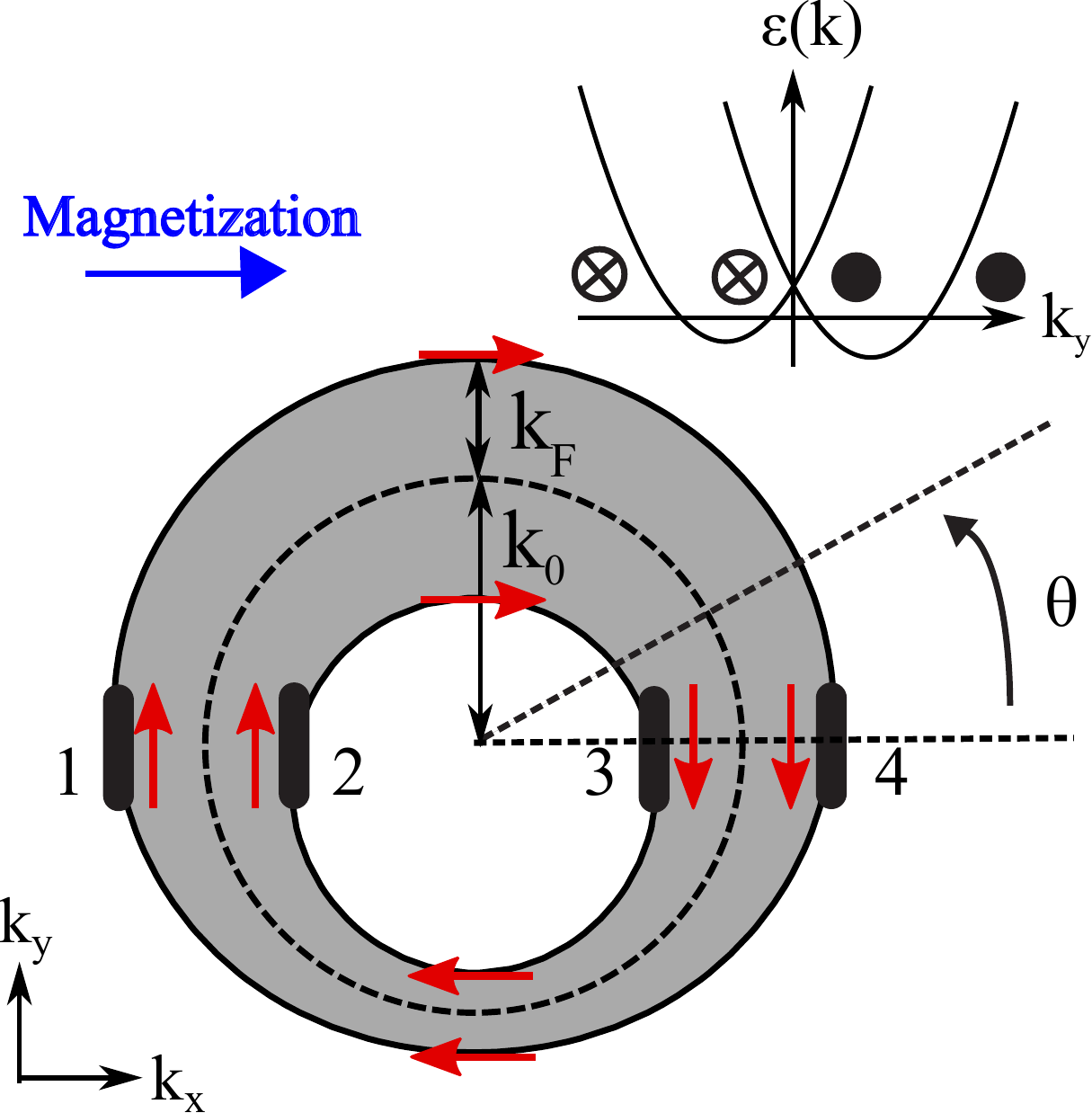} 
\hspace{.2in}
(b)\includegraphics[width=2in]{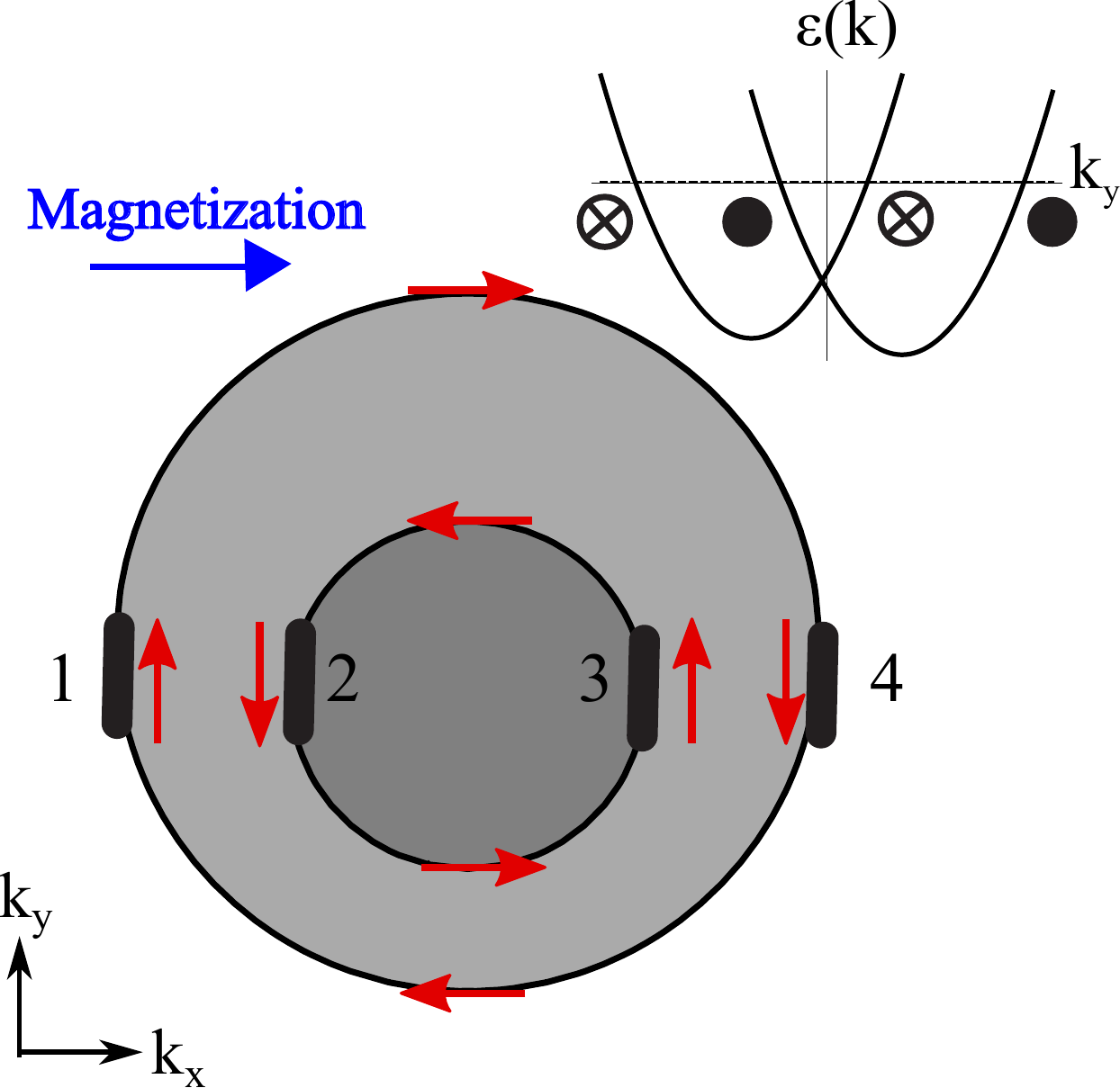} 
\caption{(a) Annular Fermi surface, obtained in the lower doping regime. Clockwise winding arrows denote electron spin orientation, which is the same on both surfaces. A representative set of collinear patches 1-4 is shown at special angle $\theta =0$ ($\theta$ measured from the x-axis), where the coupling to the Goldstone mode is strongest. The dashed circle is of radius $\ko$, while the annulus half-width is $\kf(\theta)$. In this case, the Fermi momenta are $k_F^{>,<} \equiv \ko \pm \kf(\theta)$. A cross-section of the band dispersions at angle $\theta = \pi/2, 3\pi/2$ is shown, with the directions of the electron spin pointing into (cross) or out (black) of the page. The horizontal line is the Fermi level. (b) Similar diagram for the concentric Fermi surface regime, obtained at higher doping. Note that inner, outer Fermi surfaces have opposite windings of the electron spin. The Fermi momenta are $k_F^{>,<} \equiv \kf(\theta) \pm \ko$.}
\label{fig:AnnularFS}
\end{figure*} 
\end{center}

As remarked above, in a SOC system, the spin waves are strongly coupled to low-energy electrons. The electron-spin wave interaction, Eq.~\ref{eq:Coupling}, is most conveniently treated by decomposing the Fermi surface of the non-interacting terms, $H_0\equiv H_\text{e}+H^{0}_{M}$, into patches.  Magnons with wave vector $\v{q}$ couple most strongly to patches of the Fermi surface for which $\v{q}$ is tangent since the electron dispersion is softest along this direction. Consequently, non-collinear patches decouple at low energies, and we may consider each set of collinear patches separately. Describing a (2+1)D Fermi surface in terms of decoupled (2+1)D patch theories has previously been shown to be a low-energy description that captures many physical properties. \cite{MaxIsingI,Polchinski,AIMillis,YBKim,SSLee1,SSLee2} For the TI surface, the Fermi surface decouples into pairs of antipodal patches except precisely at the Dirac point. For the 2D Rashba liquid, the FS generically has quartets of collinear patches (Fig. \ref{fig:AnnularFS}).

In imaginary time, the bare (uncorrected) linearized Lagrangian reads:
\begin{align} &\mathcal{L}_\text{patch} = \sum_j \psi^\dagger_{j,\omega,\v{k}} \(-i\omega+v_j k_\perp+\frac{k_\p^2}{2m_j}\)\psi_{j,\omega,\v{k}} \ldots \nonumber\\
&+\( \omega^{2} + c^{2}q^{2} \) |\phi_{\omega,\v{q}}|^{2} +
\sum_{j,\Omega,\v{q}} \lambda(\bs{\hat{n}_j}) \phi_{\Omega,\v{q}}\psi^\dagger_{j,\omega+\Omega,\v{k}+\v{q}}\psi_{j,\omega,\v{k}}
\label{eq:LPatch}
\end{align}
where $\bs{\hat{n}_j}$ is a unit vector in the direction of patch $j$ and $\ppp,\pp$ are coordinates perpendicular, parallel to the Fermi surface and fixed for a set of collinear patches. For the 2D Rashba liquid, $\lambda(\bs{\hat{n}_j}) = \pm \lambda_0 M_0 (\bs{\hat{e}_M}\cdot\bs{\hat{n}_j}) X_j$. Here, $X_j \equiv \frac{\alpha_R k_{F,j}}{\sqrt{ \( \alpha_R k_{F,j} \bs{\hat{z}} \cdot (\bs{\EM} \times\bs{\hat{n}_j}) + \lambda_0 M_0 \)^{2} + \( \alpha_R k_{F,j} \bs{\EM} \cdot \bs{\hat{n}_j} \)^2}} \leq 1$, $k_{F,j}$ is the Fermi momentum of the $j^\text{th}$ patch, and $\pm$ is for the lower/upper Rashba bands. For the TI, $\lambda(\bs{\hat{n}_j}) = \mp \lambda_0 M_0 \bs{\EM} \cdot \bs{\hat{n}_j}$ for the electron/hole-doped Fermi surfaces.  For strong spin-orbit coupling $\lambda_0 M_0 \ll \alpha_R k_{F,j}$, the Rashba and TI couplings are approximately the same, but for weak spin-orbit coupling $\lambda_0 M_0 \gg \alpha_R k_{F,j}$, the Rashba coupling is suppressed by a factor of $X_j \lesssim \frac{\alpha_R k_{F,j}}{\lambda_0 M_0}$ compared to the TI case.

The effect of order parameter fluctuations can essentially be treated in the random phase approximation (RPA).\cite{RPA} Deviations from this appear at higher order in the $(N, \epsilon)$ expansion. Scattering of spin waves from particle-hole excitations produces a spin wave propagator with Landau damped form:\cite{Supplement}
\begin{align}
D(\Omega,\v{q}) \approx \[\gamma(\bs{\hat{q}})\frac{|\Omega|}{|q_{\p}|}+ c^{2} q^2\]^{-1} \end{align}

\noindent which is valid when $q \gg \Omega/v_{F}$. We have dropped the bare $\Omega^{2}$ term, since at energies below $E_\text{LD} \approx \sqrt{\gamma(\bs{\hat{q}}) v_F}$ the Landau damping term dominates the dynamics of the order parameter fluctuations, leading to overdamped spin waves with dynamic exponent $z_b=3$ ($\Omega \sim q^3$).
The damping coefficient $\gamma$ depends on the coupling of a collinear set of patches at angle $\theta$ to the Goldstone mode, $\gamma \sim \lambda(\bs{\hat{n}_j})^{2}$, which for the case of strong spin-orbit coupling yields:
\begin{align} \gamma(\theta) \approx \sum_j\frac{\lambda_0^2 |m_j|}{4\pi|v_j|}\cos^2\theta. \label{eq:LDCoefficient}\end{align}
$\gamma$ is nonzero over most of the Fermi surface except for the isolated points where the local patch spin is parallel to the ordered magnetization and couples only at quadratic order to the fluctuations. The spin waves are maximally coupled to patches whose normals lie parallel/anti-parallel to the magnetic ordering direction ($\theta = 0$).  

In turn, the overdamped fluctuations scatter electrons at energy $\omega$ from the Fermi surface with rate $\Gamma \approx |\omega|^{2/3}$, corresponding to a fermion self-energy:\cite{Supplement,Polchinski,Chubukov,AIMillis,YBKim}
\begin{align}\Sigma_{f,j}(\omega)\approx i \text{ sgn}(\omega) E_\text{NFL,j}^{1/3}|\omega|^{2/3}\end{align}

Coherent Landau quasiparticles require $\frac{\Gamma}{\omega}\rightarrow 0$ as $\omega\rightarrow 0$; this signals a breakdown of these quasiparticles and of Fermi liquid theory. The NFL behavior takes over at energies below the characteristic scale:
\begin{align} 
E_\text{NFL,j} \approx \frac{\lambda(\bs{\hat{n}_j})^6}{v^3_j\gamma(\theta)c^4} \sim \frac{\lambda_0^{4}\cos^4\theta }{\ko^{2} c^4} \frac{m_0}{x \left[ 1+x + |1-x| \right]} 
\label{eq:ENFL}
\end{align}

\noindent where $x \equiv \kf/\ko \sim \sqrt{\mu}$ (see appendices or Fig. \ref{fig:AnnularFS} for definitions) and we have substituted for $\theta =0$ as an example. At high dopings, $E_{\text{NFL}}$ vanishes as $\mu^{-1}$. In contrast, the Landau damping scale $E_{\text{LD}} \sim \lambda_0 \sqrt{\frac{m_0}{x}} \left[x+1 + |x-1| \right]^{\frac{1}{2}}$ (at $\theta = 0)$ and hence approaches a constant at high doping.

\subsection{Dirac Point} 
Thus far, we have assumed a finite density of carriers. This holds everywhere except precisely at the Dirac point of the TI surface. At this fine-tuned point, the magnetic fluctuations couple to the Dirac point like a single vector component of a fluctuating $U(1)$ gauge field, and an emergent Lorentz symmetry dictates a low energy phase analogous to (2+1)d-QED.  Here the spin waves are critically damped by the Dirac fermions (see e.g. Ref. \citenum{IoffeLarkin}) and retain their relativistic dispersion (i.e. have dynamic exponent $z_b=1$ rather than $z_b=3$ characteristic of a Landau overdamped boson coupled to a finite density Fermi surface).

A fine-tuned Dirac point also arises for Rashba bands between the annular and concentric Fermi surface regimes.  However, here the Dirac point is necessarily accompanied by a large Fermi surface that Landau damps the spin waves and leads to non-relativistic $z_b=3$ scaling.


\section{Experimental considerations and candidate materials}
Having explained the basic phenomenology of the NFL phase, we now discuss two potential issues for observing the predicted NFL physics in real materials and indicate some potentially promising candidate materials.

\subsection{Crystal Anisotropy}
The assumption of continuous rotation symmetry crucial to the presence of Goldstone modes in this work is clearly broken explicitly in crystalline systems, where only discrete rotational symmetry remains. Therefore, we expect such NFL physics to be present at energy scales above that set by the crystal pinning scale, below which the Goldstone mode acquires a mass.  

It is therefore desirable to minimize the effects of crystal anisotropies.  To this end, materials with three- or six-fold rotation symmetries are preferable to those with four-fold rotation symmetry.  For the former, crystal anistropies enter at order $\mathcal{O}(k_0a)^3$, where $a$ is the lattice spacing, whereas the latter permit $\mathcal{O}(k_0a)^2$ cubic anistropy terms. Therefore, in the low density regime ($k_Fa\ll 1$), three- or six-fold anistropies permit a larger parametric separation between the spin-orbit scale $\mathcal{O}(k_0a)^2$ and the crystal pinning scale $\mathcal{O}(k_0a)^3$.  

\subsection{Disorder Effects} 
A second practical consideration is minimizing disorder.  The effects of disorder are twofold. Firstly, at temperatures less than the elastic scattering rate $\tau^{-1}$, elastic scattering of electrons from impurities dominates the inelastic scattering from spin waves and the electrons obey diffusive dynamics.  

Secondly, due to the spin-orbit coupling, impurities couple to the ferromagnetic spin texture as a random field.  Familiar Imry-Ma arguments show that this leads to random pinning of the magnetic order parameter for temperatures below some characteristic energy scale $E_\text{IM}$.  Interestingly, if $\tau^{-1}\gg E_\text{IM}$, then there will still be a broad range where the spin waves are overdamped by the diffusive fermions with damping rate $\sim q^2$ (rather than damping rate $\sim q^3$ characteristic of Landau damping\cite{HLR}).

While disorder inevitably spoils NFL physics at asymptotically low temperatures, for sufficiently clean systems there can be a broad intermediate temperature range $E_\text{NFL}\gg T \gg \tau^{-1},E_\text{IM}$ over which the NFL physics described above may be observed.

\subsection{Candidate Materials}
Promising materials with strong Rashba spin-orbit coupling and six-fold rotation symmetry include surface alloys, such as Bi/Ag(111).\cite{Ast2007,Bihlmayer2007,Ast2008,Meier2008} A complicating detail is that these surface alloys contain not only 2D SOC surface states but also 3D bulk metallic states.  The surface states inhabit regions of the surface Brillouin zone unoccupied by bulk metallic states.  However we do not expect any important modifications from the accompanying bulk states.  First, the bulk states lack strong spatial (as opposed to atomic) spin-orbit coupling and hence dynamically decouple from the spin waves as for more conventional Goldstone mode problems.  Moreover, the coupling of surface to bulk states is irrelevant for the NFL phase since the NFL physics arises from singular small momentum transfer scattering between electrons and spin waves, which cannot connect bulk and surface states which are widely separated in momentum space.

Various semiconductor heterostructures\cite{Lommer1988,Luo1990} may also be promising due to their high mobilities.  We note in passing that materials with Dresselhaus spin-orbit coupling $H_{D} \sim \alpha_{D} (\sigma_{x} k_{x} - \sigma_{y} k_{y})$ and a spin-spin coupling between electrons and an XY magnetic order parameter $\bs{M}$ are expected to give rise to similar physics. The Dresselhaus coupling preserves a continuous rotation-like symmetry of combined in-plane rotations $\mathcal{R}(\theta)$ on $\bs{M}, \bs{\sigma}$ and $\mathcal{R}(-\theta)=\mathcal{R}^{T}(\theta)$ on $\bs{k}$ which is spontaneously broken in the ground state, giving rise to a rotational Goldstone mode.  However, the combination of both Dresselhaus and Rashba spin-orbit couplings inevitably breaks the continuous rotation symmetry and does not lead to NFL behavior.

Topological insulator (TI) surface states are also promising candidates.  Since the TI surface states lack a natural tendency towards spontaneous FM, ferromagnetism may be induced by proximity in heterostructure devices between TIs and a ferromagnetic insulator\cite{FMTI} (e.g. EuO\cite{OriginalEuO} or EuS).  Ferromagnetism can also be induced in TIs by magnetic dopants; here, however, care must be taken to minimize the detrimental effects of disorder which obscure the NFL phase (as described below).

Having identified some promising experimental candidates to observe NFL behavior, we now describe how the NFL phase may be experimentally detected.

\section{Experimental Phenomenology}
\subsection{Thermodynamic Signatures}
In the NFL regime, the specific heat exhibits an unusual power law temperature dependence: $C_v\sim T^{1/z_f} \sim T^{2/3}$, which follows directly from the scaling properties of the NFL phase (see e.g. Ref. \citenum{SenthilMott} and references therein).  This quantity may be difficult to measure for a non-layered two dimensional electron system as the electronic $C_v$ is likely dominated by bulk contributions. Therefore, in subsequent sections we describe non-thermodynamic probes based on tunneling spectroscopy and electrical transport that may be more experimentally accessible.

\subsection{Spectroscopic Probes}
The characteristic non-Fermi liquid scattering rate can be directly detected by measuring the frequency dependence of linewidths in angle-resolved photoemission.  Also, the overdamped character of the spin waves may also be observable in inelastic spectroscopy.  As for heat capacity, neutron scattering is not feasible for non-layered 2D samples, but Raman spectroscopy could be used. 

Additionally, repeated scattering between electrons and spin waves produces a singular correction to the tunneling density-of-states (DOS)\cite{MaxIsingI,NEpsilon} $N(\omega) \sim |\omega|^{\eta_{f}/z_{f}}$
that could be observed in tunneling experiments or by photoemission.  Here, $\eta_f$ is an anomalous correction to the electron operator scaling dimension which appears at three and higher loop order in field theory calculations.\cite{MaxIsingI,NEpsilon} 

\subsection{Electrical Transport}
The NFL phase described above is also expected to exhibit an unusual power in the temperature dependence of the electrical resistivity.  Early studies\cite{LeeNagaosa,YBKimTransport,LeeLee,NaveLee} predicted $\rho(T) \sim T^{4/3}$ based on the scattering rate of electrons by overdamped bosons.  However this answer is likely incorrect, as momentum transferred between electrons to spin waves is not necessarily dissipated.\cite{Hartnoll} Rather, in a clean system the momentum transferred to the spin waves eventually returns to the electrons due to drag effects leading to vanishing resitivity.  Nonzero resistivity develops only from translation symmetry breaking due to impurities or umklapp scattering (the latter is typically unimportant at low temperatures\cite{Hartnoll,GiamarchiUmklapp}).

\begin{figure}[tb]
\includegraphics[width = 1.7in]{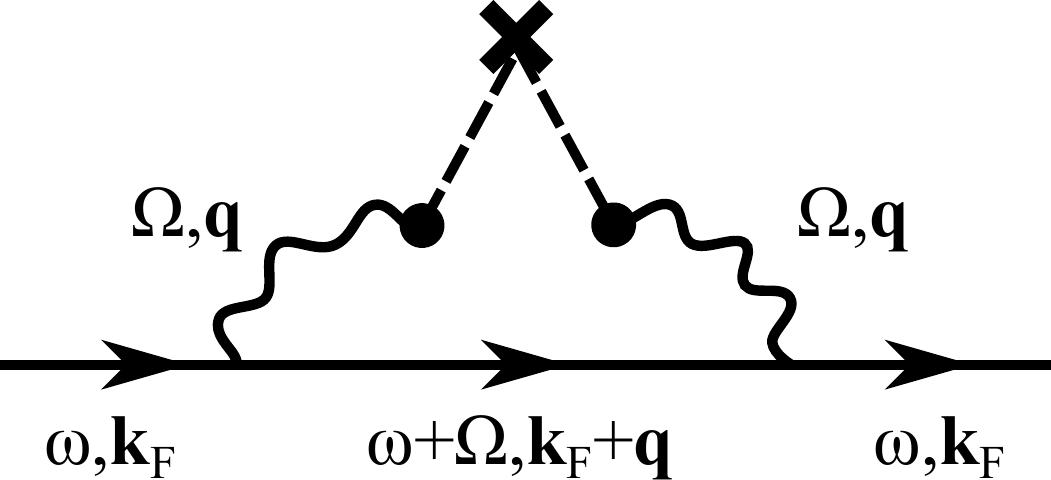}
\caption{Diagrammatic representation of the indirect disorder scattering process that is expected to set the dominant non-Fermi liquid correction to the temperature dependence of electrical resistivity.  Here straight lines indicate electron propagators, wavy lines are spin waves and dashed lines terminating at a cross indicate disorder vertices (the connection between two such lines arises from disorder averaging).}
\label{fig:DisorderDiagram}
\end{figure}

Transport for a related NFL nematic quantum critical point (QCP) in a metal was recently investigated in Ref.~\onlinecite{Hartnoll} using memory matrix techniques.  There, it was found that the dominant source of temperature dependent resistivity came from indirect momentum loss of the nematic order parameter to impurities.  Here we expect similar physics to hold in the NFL regime.  One important difference is that whereas the nematic order parameter develops a thermal mass at finite temperature, the spin waves in our problem may not develop a thermal mass at least for temperatures below the Kosterlitz-Thouless transition. As described in Appendix \ref{sec:Transportsec}, impurities couple to the spin waves as a random field.  As shown in Ref.\citenum{Hartnoll}, the results of the more sophisticated memory matrix computation can be reproduced by computing a simple low-order process in which electrons scatter from spin waves, which subsequently lose momentum to an impurity corresponding to the diagram shown in Fig.~\ref{fig:DisorderDiagram}.  For the present case, this diagram predicts resistivity that scales like:
\begin{align}\rho_\text{NFL}(T)\sim T^{2/3}\end{align}

This contribution to resistivity coexists with a constant contribution and other more conventional temperature dependent contributions from phonon scattering ($\sim T^5$) and short-range electron-electron interactions via screened Coloumb potential ($\sim T^2$).  These contributions may be distinguished from those due to spin wave scattering by applying an in-plane magnetic field.  This pins the ferromagnetic order, gaps the spin waves, and thereby suppresses the NFL resistivity contributions.  By contrast, the phonon and electron-electron scattering contributions are expected to have only weak field dependence.  Hence by comparing resistivity with and without a field one can extract the NFL contribution.

\section{Stability}
The NFL due to strong coupling between spin waves and a spin-orbit coupled FS described above bears many similarities to the nematic non-Fermi liquid proposed by Oganesyan, et al. \cite{Oganesyan}  However, subsequent analysis\cite{MaxPairing} showed that a FS coupled to gapless nematic fluctuations is strongly unstable to superconductivity. While this is a potentially interesting mechanism for high-temperature superconductivity,\cite{MaxIsingII,MaxPairing,KivelsonNematic} superconductivity was predicted to preempt the onset of NFL behavior,\cite{MaxPairing} obscuring the NFL regime underneath a superconducting dome.  


In the nematic metal, superconductivity arose since the overdamped nematic Goldstone modes mediate strong attraction between electrons on opposite sides of the Fermi surface, forming the ``glue" for Cooper pairs. For the Rashba liquid, the ferromagnetic Goldstone mode mediates attractive (repulsive) interactions for fermions with the same (opposite) spin. BCS-type superconductivity does not occur in the SOC FM metal discussed here for two reasons: 1) the spin wave mediated interaction between opposite sides of the Fermi surface is repulsive and 2) the magnetically ordered phase breaks time-reversal (TR) symmetry, energetically penalizing Cooper pairs with zero center-of-mass momentum. Spin-triplet Cooper pairing with finite center-of-mass momentum could be favored by the overdamped Goldstone modes. Ordering in a particle-hole channel, for example spin density wave order with wave vector $2k_{F}^{>}$ or $2k_{F}^{<}$, due to boson-mediated repulsive interactions is also a possibility.

Any such nonzero wave vector ordering, such as spin or pair density wave, would ultimately be fatal to the NFL phase since in the presence of translation symmetry breaking order, the rotational Goldstone modes are not independent of translational Goldstone modes. The latter couple only weakly to the Fermi surface, resulting in ordinary Fermi liquid behavior,\cite{Haruki} even if the resulting phonons have soft non-relativistic dispersions (as is the case for uniaxial density waves). Hence, whether the overdamped spin wave interactions necessarily facilitate instability to translation symmetry breaking order is hence a crucial issue for the stability of the NFL phase.  

In the remainder of this section, we show, within the combined $(N$,$\epsilon)$ expansion of Ref.\citenum{NEpsilon}, that while the susceptibilities to finite momentum pairing and spin density wave orders are enhanced compared to a Fermi liquid, they remain finite even extrapolating to the case of interest, $N, \epsilon=1$, although this lies outside of the regime of control. The computations closely follow those of the $2k_F$ susceptibilites in Ref.\citenum{NEpsilon}, but additional complications arise due to the multiple Fermi surfaces of the SOC metal.


\subsection{Validity of collinear patch theory}
The NFL theory is naturally and conveniently constructed by dividing the Fermi surface into distinct collinear patches and associating each set of collinear patches with bosonic modes having momentum tangent to the patch.\cite{NEpsilon,MaxPairing}  

However, for the nematic metal, superconductivity arises from nested scattering of Cooper pairs with zero center of mass momentum between different patches. Hence, the resulting superconducting instability is a property of the full Fermi surface and cannot be reliably obtained within a theory of decoupled patches.  This led Ref.\citenum{MaxPairing} to develop a hybrid momentum shell and patch coarse-graining renormalization group scheme ala the one developed by Son to analyze color superconductivity in non-Abelian gauge theories.\cite{Son}

By contrast, in the ferromagnetic NFL scenario discussed here, the potential instabilities all have large nonzero wave vectors.  The nonzero wave vector orders connect particle-particle or particle-hole pairs near the Fermi surface only in the vicinity of collinear patches lying along the ordering wave vector.  For example, in the case of Cooper pairing with center of mass momentum $Q=k_F^>+k_F^<$, Cooper pairs with a particle each in patches $1$ and $2$ (Fig. \ref{fig:AnnularFS}) have low energy but cannot remain close to the Fermi surface when scattered to other patches.  Hence, the susceptibility to finite momentum order is a property of the patches connected by $Q$, not the full Fermi surface, and can be computed directly within the patch theory. 

Then, due to the breaking of time-reversal and inversion symmetries, we find that the overdamped spin wave fluctuations lead to a mild enhancement of finite momentum susceptibilities but do not drive an instability towards order. 

\subsection{Patch susceptibility within the $(N,\epsilon)$ expansion}
In this section we outline the calculation, details of which are given in Appendix \ref{sec:Calcs}.
For technical computations we rely on the $(N,\epsilon)$ expansion technique.\cite{NEpsilon} In this approach the problem is generalized to have $N$ flavors of fermions and modified bosonic kinetic energy $|q|^{1+\eps}|\phi_{\omega,\bs{q}}|^2$ (instead of the physical $q^2|\phi_{\omega,\bs{q}}|^2$ of Eq.~\ref{eq:LPatch}).  The expansion is justifed in the joint limit of large $N$ and small $\epsilon$ while maintaining the product $N\epsilon\sim \mathcal{O}(1)$.  As a cautionary note, obtaining results for the physically relevant values $N, \epsilon=1$ requires extrapolation beyond the safely controlled regime (as is typical for such asymptotic expansions).  
However, in lieu of further theoretical developments, the $(N,\epsilon)$ expansion is essentially the current state of the art. 

\begin{center}
\begin{figure}[t]
(a)\includegraphics[width=2.5in]{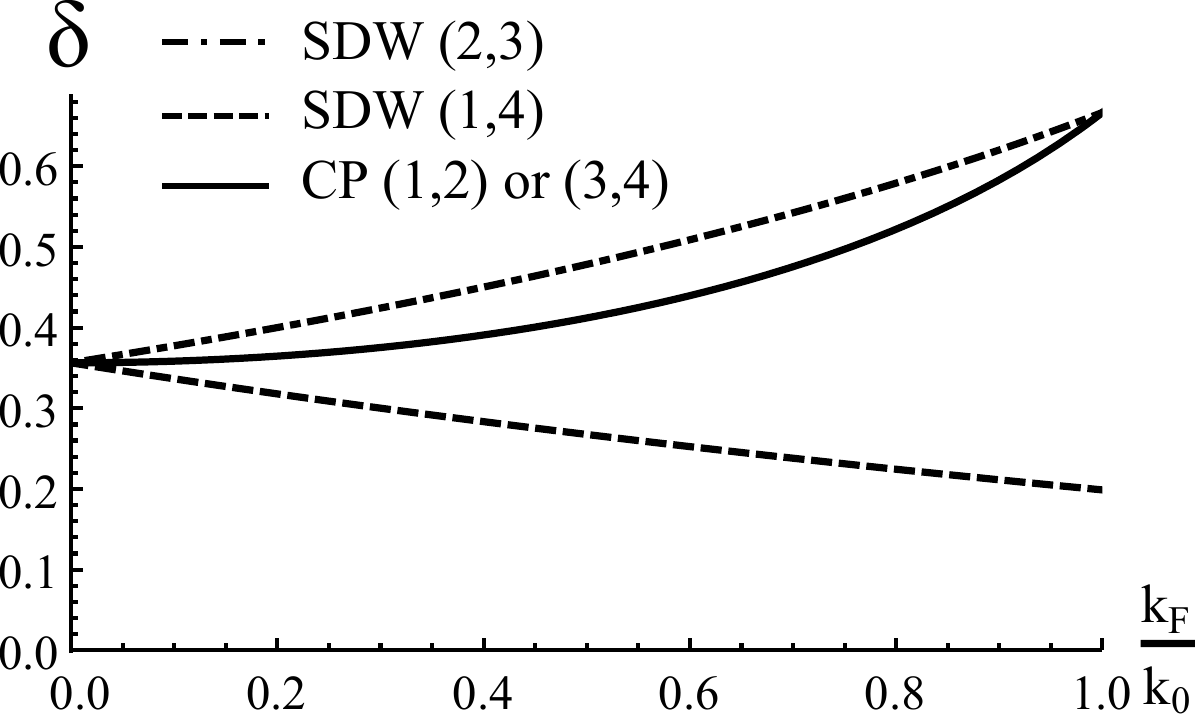} 
\\
\hspace{.2in}
\\
(b)\includegraphics[width=2.5in]{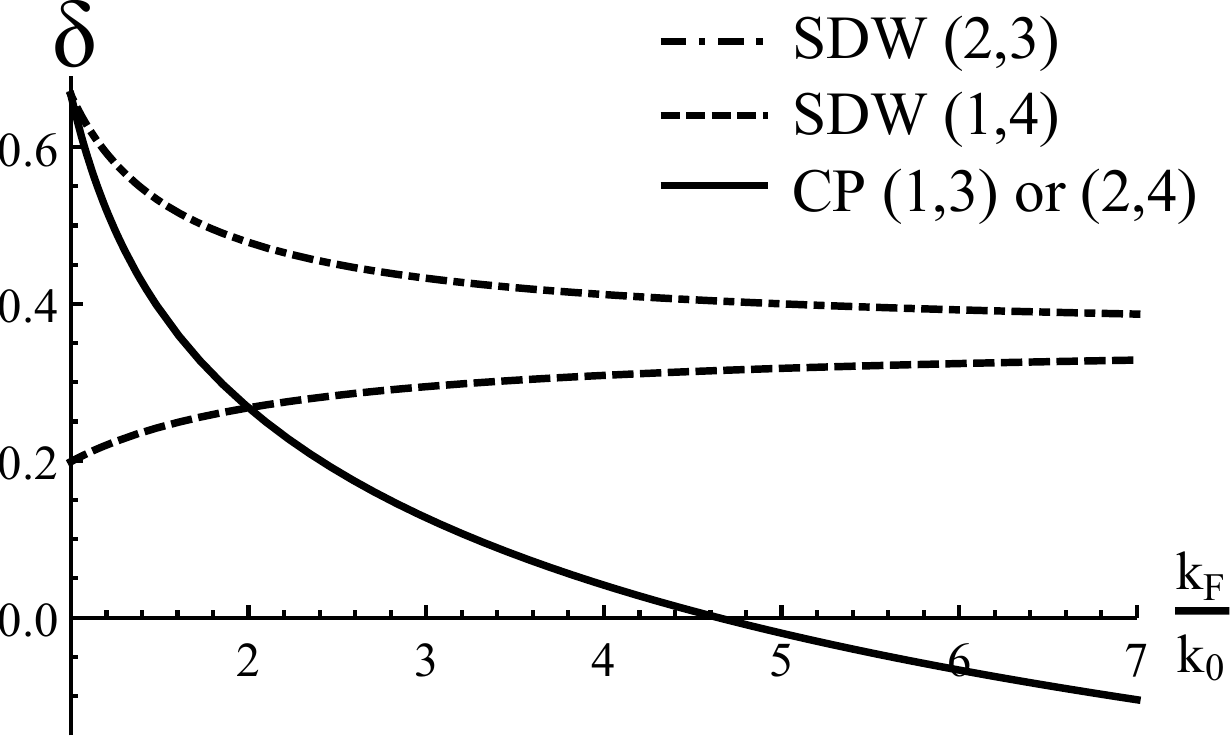} 
\caption{Scaling exponents $\delta_a$ for susceptibilities $\chi_a(\omega) \sim |\omega|^{\delta_a}$ of various orders, $a$, that are enhanced by the interaction between electrons and overdamped spin waves. CP/SDW $(j,j')$ denotes Cooper pairing or spin density wave between patches $j$ and $j'$ (see Fig.~\ref{fig:AnnularFS} for patch labeling convention). Results are obtained by continuing the large $N$, small $\epsilon$ expansion to the physical values $N,\epsilon=1$ and are shown as a function of electron density, parameterized by the ratio $\frac{k_F}{k_0}$. (a) Lower doping of the Rashba liquid (annular FS), when $0 < \frac{k_F}{k_0} < 1$. 
Note, in this case the susceptibilities are mildly enhanced but no instability develops. Curves correspond to Cooper pairing between either (1,2) or (3,4) (solid), SDW order between (1,4) (dashed), and SDW order between (2,3) (dot-dashed). (b) Concentric FS, given by $\frac{k_F}{k_0} > 1$. Curves correspond to Cooper pairing between either (1,3) or (2,4) (solid), SDW order between (1,4) (dashed), and SDW order between (2,3) (dot-dashed). At high densities $k_F\gg k_0$, nesting in the Cooper channel is approximately restored by the strong spin-orbit coupling and a superconducting instability develops (indicated by $\gamma_\text{CP}<0$).}
\label{fig:SpecificNeps}
\end{figure}
\end{center}

Another potentially promising alternative is the recently developed codimension expansion.\cite{SSLeeCodim,SSLeeChiral} Despite some potential advantages, the codimension expansion clearly overestimates the stability of a Fermi surface to superconducting order (for familiar Fermi liquid interactions the codimension suppresses the well-known BCS instability towards superconductivity). For this reason we believe the $(N,\epsilon)$ expansion better suited for our analysis of stability.  Given the complexity of this strongly coupled problem, experiments which can test the validity of various theoretical approaches are highly desirable.

We evaluate the system's response to a test field of strength $u_a$ that couples to an order parameter $\mathcal{O}_a(\bs{x},\tau)$ in channel $a$. For instance, for $a = 2 k^>_F$ SDW, a term $S_{\text{ext}} = \int d^{2}\bs{x} d\tau \left[u_{2 k^>_F} \psi^{\dag}_{1} \psi_{4} (\bs{x},\tau) + h.c. \right]$ would be added to the action. We are interested in the scaling form of two-point correlations of these order parameters. This can be deduced from the scaling dimension of the test field $u'_a = b^{\phi_{u_a}} u_a$, for which we compute the leading correction in $\epsilon$ (or equivalently, $N^{-1}$).\cite{Supplement} Defining $\chi_{a}$ as the Fourier transform of the two-point correlation of $\mathcal{O}_{a}$, namely $\langle \mathcal{O}^{*}_{a}(x_{\perp},x_{||},\tau) \mathcal{O}_{a}(0,0,0) \rangle$, it obeys the scaling form:

\be
\bsplit
\chi_{a}(\ppp, \pp,\omega)= |\omega|^{\delta_a} F_a\left[ \frac{|\omega|}{|\pp|^{\zb}}, \frac{\ppp}{\pp^{2}} \right]\\
\end{split}
\ee

\noindent with $z_b$ the boson dynamic critical exponent, the power $\delta_a = 1 + (3 - 4\phi_{u_a})/\zb$, and $F_a$ is a scaling function. For the case of physical interest ($N, \epsilon = 1$) and taking the angle $\theta=0$ (where the electron-boson coupling is strongest), we find:

\be
\bsplit
\delta_a &= \frac{2}{3}\left[1 - \frac{g(\sqrt{3}R_a/\pi,\zb=3)}{R_a} \right]
\end{split}
\ee
where the functional form of $g(x,\zb)$ is in Appendix \ref{sec:Calcs} and is independent of the susceptibility channel. $R_a$ is a dimensionless ratio weighing the mass contribution in the Landau damping $\gamma$, which originates from both inner and outer Fermi surfaces, against a channel ($a$) dependent ``effective" DOS of the two patches under consideration. $\delta_a$ is a monotonically increasing function of $R_a$. Fig. \ref{fig:SpecificNeps} shows the value of the power $\delta_a$ for the four channels of interest as a function of the more tunable parameter $k_{F}/\ko$ which increases monotonically with chemical potential, $k_{F}/\ko \sim \sqrt{\mu}/(\sqrt{\mo}\alpha_R)$. The susceptibilities in the particle-hole channel (at all dopings) and Cooper channel (annular FS, i.e. low dopings) remain finite. Only in the case of Cooper pairing at higher dopings (concentric FS) is there potential for a singularity in the susceptibility. In this case, when $\kf >> \ko$ one asymptotically restores the time-reversal symmetry nested Fermi surfaces that necessitate the use of a full Fermi surface RG scheme as in Ref. \citenum{MaxPairing}. 

Lastly, we remark that for sufficiently low densities and strong magnetization $M_0$, the Fermi surface has only a single ``banana" shaped pocket centered at non-zero momentum perpendicular to the magnetic ordering direction (see e.g. \citenum{Berg,RuhmanBerg}).  In this regime only the Cooper pairing $(1,2)$-type channels are available, and our analysis again suggests that this regime is also a stable non-Fermi liquid.

\section{Conclusion}
We have shown that Fermi liquid theory breaks down in spin-orbit coupled metallic ferromagnets with broken continuous rotational symmetry\cite{CenkeTIFM} and have highlighted experimentally testable signatures of this non-Fermi liquid phase.  Importantly, we find that this non-Fermi liquid (NFL) phase is stable to subsidiary symmetry breaking, which would disrupt the NFL behavior.  Unlike the related nematic metal problem, the spin-orbit coupled metallic ferromagnets do not suffer an instability towards superconductivity due to the absence of time-reversal and inversion symmetries.  We also analyzed instabilities in other pairing and spin density wave channels within a controlled $\(N,\epsilon\)$ expansion.\cite{NEpsilon}  While certain susceptibilities receive non-analytic enhancements from strong spin wave mediated interactions, we find that no instabilities develop over a wide range of carrier densities.  

The realization of this NFL phase is likely experimentally feasible.  Promising candidate materials include surface alloys, topological insulator surface states, and semiconductor heterostructures.  While related non-Fermi liquids are expected to arise in more exotic contexts like quantum critical points in metals or gapless spin liquids with emergent gauge fields, spin-orbit coupled metallic ferromagnets offer an experimentally accessible and comparatively simple platform for exploring the physics of correlated gapless quantum phases without quasiparticles.\\

\noindent\textit{Acknowledgements - }
We thank H. Watanabe, A. Vishwanath, M. Metlitski, S.-S. Lee, T. Senthil, J. Ruhman, and E. Berg for helpful conversations. This work was supported by NSF GRFP under Grant No.
DGE 1106400 (YB) and the Gordon and Betty Moore Foundation (ACP).

\vspace{12pt}

\noindent\textit{Note - }
A related paper, Ref.~\onlinecite{RuhmanBerg}, appeared during the completion of this manuscript.  This work builds on Ref.~\onlinecite{Berg} to analyze the potential magnetic instabilities of the low-density Rashba coupled electron liquid (encouragingly finding a regime where ferromagnetism is favored at low densities) and notes the emergence of non-Fermi liquid behavior in the FM phase originally identified in Ref.~\onlinecite{CenkeTIFM}.

\appendix 

\section{Temperature dependence of electrical resistivity in the NFL phase}
\label{sec:Transportsec}
In this Appendix, we introduce a model for the impurity/spin wave coupling and include details of the computation of the temperature dependence of electrical resistivity.

\subsection{Disorder model}
Nonmagnetic impurities produce a spatially random potential: $\hat{V}=\int d^2 \bs{r} \left[\sum_\sigma \psi_\sigma^\dagger(\v{r})\psi_\sigma(\v{r}) \right] V_\text{imp}(\v{r})$. The following coupling between the impurites and the magnetization $\v{M}(\v{r})$:
\begin{align}
V_{\v{M},\text{imp}} = \lambda_{\text{imp}}\int d^2 \bs{r} \[\bs{\hat{z}}\cdot \(\v{E}_\text{imp}(\v{r})\times \v{M}(\v{r})\)\]^2
\label{eq:VMimp}
\end{align}
is symmetry allowed\cite{BergPC} and hence will be generated upon integrating out high energy electronic modes to obtain a long wavelength effective action for the magnetization.  Here $\v{E}_\text{imp} = -\nabla V_\text{imp}(\v{r})$ is the local electric field due to the impurites, and $\lambda_\text{imp}$ is proportional to the square of the Rashba spin-orbit coupling strength $\alpha_R$.

In the ferromagnetic phase, we may decompose the magnetization into ordered and spatially fluctuating pieces: $\v{M}(\v{r},t)=\v{M}_0+\delta \v{M}(\v{r},t)$.  Rotational Goldstone modes correspond to  magnitude-preserving orientational fluctuations $\phi$, which to linear order in $\phi$ can be written as $\delta\v{M}(\v{r},t)=\bs{\hat{z}}\times \v{M}_0\phi(\v{r},t)$. Eq.~\ref{eq:VMimp} becomes:
\begin{align}
V_{\v{M},\text{imp}} &\approx  \int d^2\bs{r} \left[ h(\v{r}) \phi(\v{r},t) \right]+\mathcal{O}\(\phi\)^2
\nonumber\\
h(\v{r})&=2\lambda_{\text{imp}}\[\bs{\hat{z}}\cdot \(\v{E}_\text{imp}(\v{r})\times \v{M}_0(\v{r})\)\]\[\v{E}_\text{imp}(\v{r})\cdot \v{M}_0(\v{r})\]
\label{eq:LinearCoupling}
\end{align}
Hence, we see that disorder couples linearly to the rotational Goldstone modes $\phi(\v{r},t)$.  Then, standard Imry-Ma arguments show that this linear coupling will pin the Goldstone modes in a random fashion destroying the NFL phase for temperatures below some characteristic scale $E_\text{IM}$.

For sufficiently weak disorder, however, there will be a wide range of temperatures $E_\text{IM}\ll T\ll E_\text{NFL}$ over which NFL physics may be observed.  In this regime, disorder scattering will produce nonzero electrical resistance which we now estimate.

\subsection{Temperature dependence of resistivity}
For simplicity, we treat the random field $h(\v{r})$ as being normally distributed independently for each position $\v{r}$:
\begin{align}
\overline{h(\v{r})h(\v{r}')}=h_0^2\delta^2(\bs{r}-\bs{r'})
\end{align}
where $\overline{\(\dots\)}$ indicates averaging over disorder configurations.  This approximation is reasonable for weak or dilute impurities and is expected to reproduce universal behavior, such as the temperature dependence of resistivity, for more generic impurity distributions.  We now estimate the contribution to resistivity from random impurities scattering spin waves via the linear coupling in Eq.~\ref{eq:LinearCoupling}.

It was shown in Ref.~\onlinecite{Hartnoll}, for the closely related problem of a nematic quantum critical point in a metal, that the dominant temperature dependence of resistivity comes from impurity scattering of the Landau damped bosons (in our case spin waves).  Moreover, it was shown that the results of the more sophisticated memory matrix formalism could be reproduced by perturbatively computing the effective momentum loss rate by the process equivalent to the imaginary part of the diagram shown in Fig.~\ref{fig:DisorderDiagram}, in which an incoming electron emits an overdamped boson which loses momentum to the impurites.  In this appendix, we compute the analogous diagram for the NFL described in the main text.  More sophisticated treatments, e.g. using memory matrix formalism, are left for future work.

%
%

Evaluating the diagram shown in Fig.~\ref{fig:DisorderDiagram} gives: 
\begin{align} \Sigma_\text{tr}(\Omega)&\approx \int d\Omega dq_{\perp} dq_{||} (1-\cos\theta_{\bs{q}})D(\Omega,\bs{q})^2G(\omega+\Omega,\v{q}) 
\nonumber\\
&\underbrace{\approx}_{\int dq_\perp} \int d\Omega dq_\p \left[ q_\p^2\times i\text{ sgn}(\omega+\Omega) \frac{q_\p^2}{\(|\Omega|+|q_\p|^{3}\)^2} \right]
\nonumber\\
&\approx ih_0^2~\text{sgn}{(\omega)}~|\omega|^{2/3}
\end{align}
Here, as is typical for the computation of transport scattering rates (see also Ref.~\onlinecite{Hartnoll}), we have included a factor of $(1-\cos\theta_{\bs{q}})\approx \(\frac{q_\p}{k_F}\)^2$ to appropriately weigh small angle scattering that does not substantially change the electron momentum.  Analytically continuing to retarded frequency $i\omega\rightarrow  \omega+i0^+$ and trading the low frequency cutoff $\omega$ for temperature $T$ gives the temperature dependence $\rho(T)_\text{NFL}\sim T^{2/3}$ quoted in the main text.


\section{Calculations for the NFL and its stability}
\label{sec:Calcs}

\subsection{Setup}

The Rashba Hamiltonian in the presence of a uniform magnetization $\bs{M} = M_0 \bs{\EM}$ is 

\be
\bsplit
H_{0} = \frac{k^2}{2\mo} -\tilde{\mu} + \alpha_R \bs{\zhat} \cdot (\v{k} \times \bs{\sigma} ) - \lambda_0 M_0 \bs{\EM} \cdot \bs{\sigma}
\end{split}
\ee

\noin with dispersion $E(k,\theta) = \frac{(k-\etap \ko)^{2}}{2\mo} - \mu -\etap \lambda_0 M_0 \sin(\theta)$, where $\etap = \pm 1$ denotes the lower/upper Rashba bands, $\ko \equiv \mo \alpha_R$, $\mu = \tilde{\mu} + \frac{\mo \alpha^{2}_R}{2}$, and $\theta$ is measured from the direction of $\bs{\EM}$. Using $\eta = \pm 1$ to further denote the outer/inner Fermi surface and defining $\kf(\theta) \equiv \sqrt{2\mo\left[ \mu + \etap \lambda_0 M_0 \sin(\theta) \right]}$, the radius of a FS is $\etap (\ko + \eta \kf) >0$ assuming there is a Fermi surface at the given angles and doping (e.g. not a Fermi pocket). We break the FS into patches (a representative set labeled 1-4 is shown in Fig. \ref{fig:AnnularFS}). Low-energy quasiparticles in a patch obey dispersion $\varepsilon = v \kpp + \frac{\kp^{2}}{2m}$, with $\kpp,\kp$ coordinates that are perpendicular, parallel to the FS and are fixed with respect to a set of patches; the signed velocity, effective mass are $|v| = \kf/\mo$, $ m = \mo (1 + \eta \frac{\ko}{\kf})$. We denote $k^{>,<}_{F}$ as shorthand for the radius of the outer, inner FS. 

The full fluctuating magnetization is parameterized as $\bs{M} = M_0 \left[ \bs{\EM} \cos\phi + (\bs{\zhat} \times \bs{\EM}) \sin\phi \right]$. We expand for small fluctuations $\phi$ and also redefine $M_0 \phi \rightarrow \phi$. We will denote $\lambda(\bs{\hat{n}_{j}})$ the coupling to the Goldstone mode, where $\bs{\hat{n}_{j}}$ is a unit vector in the direction of patch $j$. Expressions for the functional dependence of $\lambda(\bs{\hat{n}_{j}})$ can be found in the main text. \\

\subsubsection{Special Case: TI surface}

The Hamiltonian for the TI surface state is 
\be
\bsplit
H_{0} = \vd \bs{\zhat} \cdot (\v{k} \times \bs{\sigma}) - \lambda_0 M_0 \bs{\EM} \cdot \bs{\sigma} - \mu
\end{split}
\ee
 
If we solve $H_{0}$ exactly, the effect of the magnetization, which couples like a gauge field to the fermions, is only to shift the center of the Dirac cone. There are in general higher order in $\bs{k}$ corrections in the Hamiltonian which will explicitly break the continuous rotational symmetry of the Fermi surface, but we neglect these here. The case of the TI surface state is accounted for within the treatment of the Rashba spin-orbit coupled system. The patch dispersion for the TI is $\varepsilon = \vd \kpp + \frac{\vd^{2} \kp^{2}}{2\mu}$ where the patch velocity $v=\vd$ and mass $m = \mu/\vd^{2}$.

\subsection{Lagrangian}

The initial Euclidean action for a set of collinear patches $j$ within the $(N,\epsilon)$ expansion of Ref. \citenum{NEpsilon} is:

\be
\bsplit
S_{E} &\equiv S_{0} + S_{\text{int}} \\
S_0&= \int \frac{d\om d^{2}\bs{k}}{(2\pi)^{3}} \Big{[} N c^{\zb} |\kp|^{\zb-1} |\phi_{\om,\bs{k}}|^{2}  
\\
&+ 
\sum_{\mu,j} \psi^{\dag,\mu}_{j,\om,\bs{k}} \left( -i\om + \vj \kpp + \frac{\kp^{2}}{2\mj}\right) \psi^{\mu}_{j,\om,\bs{k}}\Big{]} + 
\\
S_\text{int}&= \sum_{\mu,j} \int \frac{d\OM d^{2}\bs{q}}{(2\pi)^{3}}\[ \lambda(\bs{\hat{n}_{j}}) \phi_{\OM,\bs{q}} \psi^{\dag \mu}_{j,\om+\OM,\bs{k} + \bs{q}} \psi^{\mu}_{j,\om,\bs{k}} \]
\label{eq:GenAction}
\end{split}
\ee

\noindent where $\mu$ indexes $N$ fermion flavors and $j$ indexes patches 1-4 (Fig. \ref{fig:AnnularFS}). (We will omit the $\mu$ flavor index where it is unnecessary to track). The control parameter $\epsilon$ is related to the boson dynamic critical exponent $z_b$ as $\epsilon \equiv z_b -2 >0$. This is expected to be useful when the exponent is not renormalized, for instance if it governs a nonlocal term in the Lagrangian.\cite{NEpsilon}

This action will generate the one-loop boson polarization and fermion self-energy computed below and together these give the non-Fermi liquid action around which we will work perturbatively. Initially, we define the free propogators as $\langle |\phi_{\om,\bs{k}}|^{2} \rangle = (2\pi)^{3} (N c^{\zb} |\kp|^{\zb-1})^{-1} \equiv (2\pi)^{3} D_{0} (\om,\bs{k}), \langle \psi^{\mu}_{j,\om,\bs{k}} \psi^{\dag \mu}_{j,\om,\bs{k}} \rangle  = (2\pi)^{3} (-i\om + \varepsilon_{j})^{-1} \equiv (2\pi)^{3} G^{j}_{0} (\om,\bs{k})$ with $\varepsilon_{j} = \vj \kpp + \frac{\kp^{2}}{2\mj}$. 

\begin{center}
\begin{figure}[ttt]
(a)\includegraphics[width=1.3in]{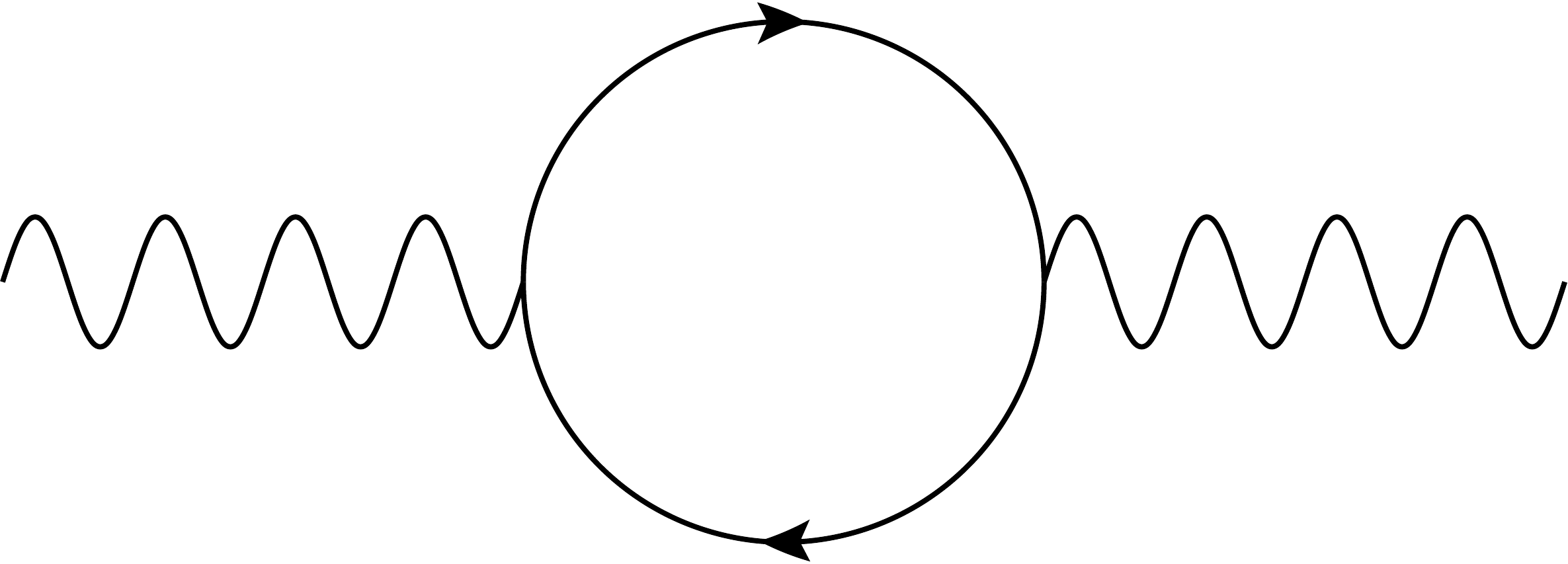}
(b) \includegraphics[width=1.3in]{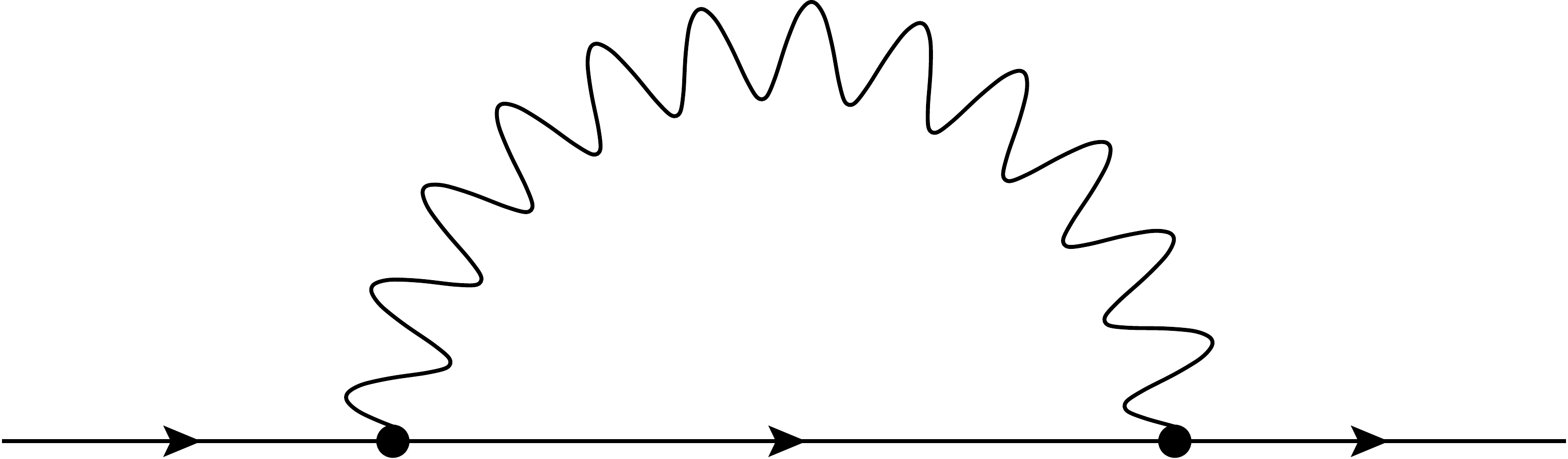}
\caption{(a) Boson polarization. (b) Fermion self-energy.}
\label{fig:BosonP}
\end{figure} 
\end{center}

\subsection{Boson polarization (1-loop)}

Defining the boson polarization via Dyson's equation as $D^{-1} = D^{-1}_{0} + \Pi$, we obtain:

\be
\bsplit
\Pi&(\OM,\bs{q}) 
\nonumber\\
&= N \sum_{j} \lam(\bs{\hat{n}_{j}})^{2} \int \frac{d\om d\kp d\kpp}{(2\pi)^{3}} G^{j}_{0}(\om,\bs{k})G^{j}_{0}(\om+\OM,\bs{k}+\bs{q})
\end{split}
\ee
Performing the integrals in the order $\kpp,\om,\kp$ gives:
\be
\bsplit
\Pi(\OM,\bs{q}) &= N \sum_{j} \lam(\bs{\hat{n}_{j}})^{2} \frac{|\mj|}{4\pi|\vj|} \frac{|\OM|}{|\QP|} 
\equiv N \gamma \frac{|\OM|}{|\QP|}
\end{split}
\ee
which yields the Landau damping coefficient $\gamma$ depending on the patch velocity, mass, and effective coupling $\lam$ to the Goldstone mode.  

\subsubsection{Landau damping energy scale} 
The boson velocity is expected to be renormalized by the fermions, via e.g. the diagram in Fig. \ref{fig:BosonP}, and hence acquire a component proportional to $v_F$. Using $\Omega \sim v_F q$, the energy scale below which damped dynamics dominates is $E_{\text{LD}} \sim (\gamma v_F)^{\frac{1}{2}}$. Considering $\theta =0$, where the masses of patches 1-4 (Fig. \ref{fig:AnnularFS}) are $|m_1|=|m_3|, |m_2|=|m_4|$, $\gamma \sim \lambda^{2} \frac{(|m_1| + |m_2|)}{v_F}$. Defining $x \equiv \kf/\ko \geq 0$, a quantity which increases monotonically with doping, we have
\be
\bsplit
E_{\text{LD}} \sim \lambda_0 \sqrt{\frac{m_0}{x}} \left[x+1 + |x-1| \right]^{\frac{1}{2}}
\end{split}
\ee
With varying chemical potential but other parameters held fixed, $k_F \sim \sqrt{\mu}$ and so $E_{\text{LD}} \sim (\mu)^{-\frac{1}{4}}$ for low doping $\mu << \mo \alpha^{2}_R$ while it approaches a constant at high doping $\mu >> \mo \alpha^{2}_R$.

\subsection{Electron self-energy (1-loop)}
Using the Landau damped Goldstone boson propagator, we compute the frequency dependence of the electron self-energy for a single patch $j$ defined by $(G^{j})^{-1} \equiv (G^{j}_{0})^{-1} - \Sigma_{j} = -i\omega + \varepsilon_{j}(\bs{k}) - \Sigma_{j}(\omega)$ at $\bs{k}=0$. Within the RPA, the momentum dependence is expected to be IR nonsingular.\cite{NEpsilon}
\be
\bsplit
\Sigma_{j}(\om,0) &= \lam(\bs{\hat{n}_{j}})^{2} \int \frac{d\OM d\QP d\QPP}{(2\pi)^{3}} D(\OM,\bs{q}) G^{j}_{0}(\om-\OM,-\bs{q})
\nonumber\\
&=i \frac{\text{sgn}(\omega)}{\zeta N} |\omega|^{2/\zb}
\end{split}
\ee
\be
\bsplit
\zeta &\equiv 4\pi \frac{|\vj|}{\lambda(\bs{\hat{n}_{j}})^{2}} \gamma^{(\zb-2)/\zb} c^{2(1-\frac{1}{\zb})} \sin(2\pi/\zb)\\
&\approx   \(\frac{2\pi^{2} c |\vj|}{\lambda(\bs{\hat{n}_{j}})^{2}} \) \epsilon \hspace{.2in}\text{  (for } \epsilon \ll 1)
\end{split}
\ee\\

\subsubsection{NFL energy scale} 
We consider the physically relevant case $\zb=3, N=1$. The scale below which the NFL sets in is estimated as $E_{\text{NFL}} = \zeta^{-3}$. With $x \equiv \kf/\ko$,
\be
\bsplit
E_{\text{NFL}} \sim \frac{\lambda^{4}_0 \cos^{4}\theta}{\ko^{2} c^{4}} \frac{\mo}{x \left[ 1+x + |1-x| \right]} 
\end{split}
\ee
Since $\kf \sim \sqrt{\mu}$, $E_{\text{NFL}} \sim 1/\sqrt{\mu}$ for  $\mu << \mo \alpha^{2}_R$ while $E_{\text{NFL}} \sim 1/\mu$ for $\mu >> \mo \alpha^{2}_R$. The non-Fermi liquid energy scale vanishes as $\mu \rightarrow \infty$.   

\begin{center}
\begin{figure}[ttt]
(a)
\includegraphics[width=1.7in]{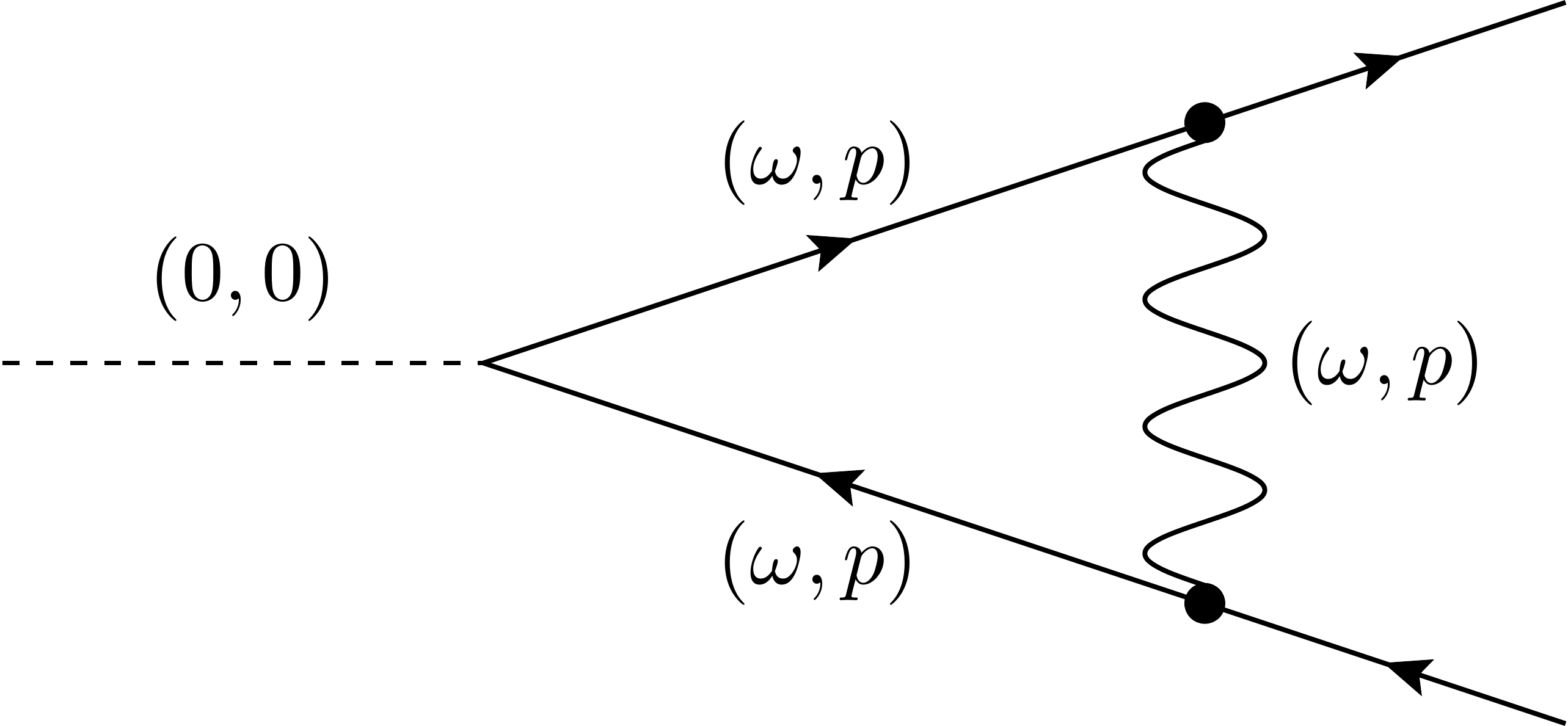} 
\\ 
\text{   }
\\
(b)
\includegraphics[width=1.7in]{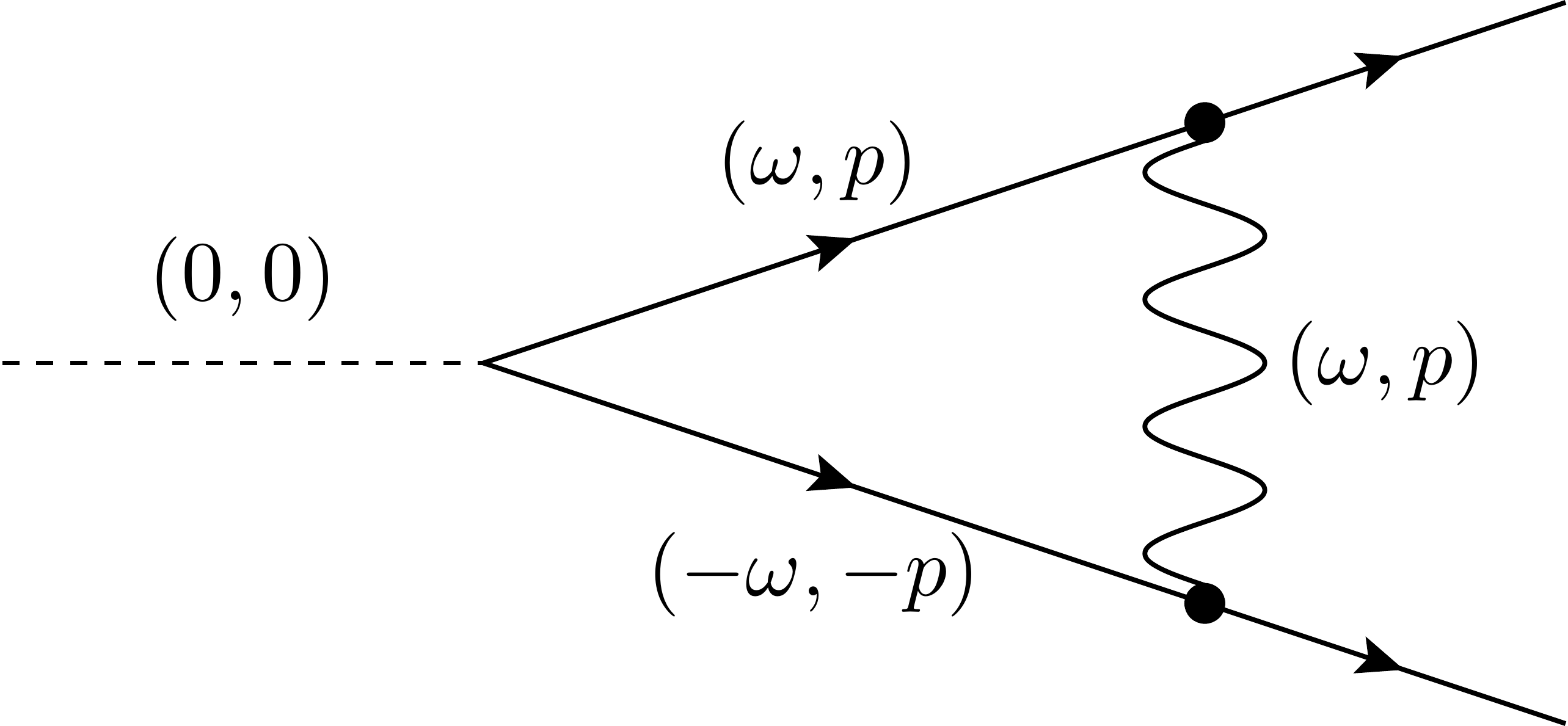} 
\caption{Vertex correction for coupling to an external field $u$ (dashed) due to the interaction between electrons (solid) and Goldstone boson (wavy). The two channels are (a) particle-hole and (b) Cooper pair.}
\label{fig:VertexFig}
\end{figure}
\end{center}

\subsection{Patch susceptibilities}
We consider the finite momentum particle-hole and Cooper pair susceptibilities for patches within a collinear set in the Rashba system. We add a term $S_{\text{ext}}$ to the action which couples an external field $u$ to electron patch bilinears and compute the renormalization of the coupling due to the electron-boson interaction. The modified scaling dimension of the coupling will yield the scaling form for the susceptibility to ordering in a channel. The leading order in $\epsilon$ (equivalently, $1/N$ since $\epsilon N \sim O(1)$) correction comes from Fig. \ref{fig:VertexFig}. 

The action consists of $S_0 + S_\text{int}+S_\text{ext}$ with $S_\text{int}$ from Eq. \ref{eq:GenAction} but  the noninteracting boson and fermion actions modified to include Landau damping and the NFL self-energy:
\be
\bsplit
S_0 &= N \int \frac{d\om d^{2}\bs{k}}{(2\pi)^{3}} \left[ c^{\zb} |\kp|^{\zb-1} + \gamma \frac{|\omega|}{|\kp|} \right] |\phi_{\om,\bs{k}}|^{2}  + \\
&+\sum_{\mu,j} \int \frac{d\omega d^{2}\bs{k}}{\Twopic} \psi^{\dag \mu}_{j}\left[ -\frac{i \text{ sgn}(\omega)}{\zeta N}|\omega|^{2/\zb} + \vj \ppp + \frac{\pp^{2}}{2\mj} \right] \psi^{\mu}_{j}
\end{split}
\ee

$S_{\text{ext}}$ has the general form $S_{\text{ext}} = \int d^{2}\bs{x} d\tau \left[ u_a \mathcal{O}_a(\bs{x},\tau) + h.c. \right]$ and depends on the channel ($a$) under consideration. For instance, $\mathcal{O}_a(\bs{x},\tau) = \psi^{\dag}_{1} \psi_{4}(\bs{x},\tau)$ for the $a=2k^>_F$ SDW channel. The RG scheme involves integrating out low-energy electron and boson modes simultaneously. The RG scaling transformations are:\cite{NEpsilon}
\be
\bsplit
\om' &= \om b^{\zb/2}\\
\ppp'&=\ppp b\\
\pp' &= \pp \sqrt{b}
\end{split}
\qquad
\qquad
\begin{split}
\psi'(\bs{k}',\om') &= b^{-(\zb+5)/4}\psi(\bs{k},\om)\\
\phi'(\bs{k}',\om') &= b^{-(1+\zb)/2} \phi(\bs{k},\om)
\end{split}
\ee
which keeps $S_{0}$ invariant. Note that the $\perp, ||$ directions rescale differently. To carry out the RG, we use $\int_{>} d\omega d\pp d\ppp$ which integrates over all $\omega,\ppp \in (-\infty,\infty)$ but over $|\pp| \in \left[ \Lambda/\sqrt{b}, \Lambda \right]$. \\

The renormalized scaling dimension of the external field $u_a$ can be related to the scaling form of the correlations. If $u'_a=b^{\phi_{u_a}} u_a $ under renormalization, the order parameter $\mathcal{O}_{a}(\bs{x},\tau)$ to which $u_a$ couples transforms as $\mathcal{O'}_{a}(\bs{x}',\tau') =  b^{\frac{(\zb + 3)}{2} - \phi_{u_a}}\mathcal{O}_{a}(\bs{x},\tau)$. We define $\chi_{a}$ as the Fourier transform of the two-point correlation $\langle \mathcal{O}^{*}_{a}(\bs{x},\tau) \mathcal{O}_{a}(0,0)\rangle$ and it obeys the relation:
\be
\bsplit
\chi_{a}(\bs{k},\omega)=  b^{2\phi_{u_a} - \frac{(\zb + 3)}{2}} \chi'_{a} (\bs{k}',\omega')
\end{split}
\ee
leading to the scaling form:
\be
\bsplit
\chi_{a}(\ppp, \pp,\omega)= |\omega|^{\delta_a} F_a\left[ \frac{|\omega|}{|\pp|^{\zb}}, \frac{\ppp}{\pp^{2}} \right]
\end{split}
\ee
with the power $\delta_a = 1 + \frac{(3 - 4\phi_{u_a})}{\zb}$ and scaling function $F_a$. 

As the Goldstone mode mediates repulsive interactions in the particle-hole channel and attractive interactions in the finite momentum Cooper channel, correlations in these channels are expected to be enhanced compared to the usual Rashba Fermi liquid and hence are of interest for potential singularities.

\subsubsection{$2k^{<,>}_F$ particle-hole susceptibility}

We compute the one-loop correction to the coupling $S_{\text{ext}} = \int d^{2}\bs{x} d\tau \left[ u_a \psi^{\dag}_{j} \psi_{j'} + h.c. \right]$ with $a=2k^>_F$ or $2k^<_F$. From this, we obtain the particle-hole susceptibility to SDW order within patch pairs $(j,j') = (1,4)$ and $(2,3)$ in a collinear set, corresponding to finite momentum $2k^>_F$ and $2k^<_F$, respectively. This channel is relevant for both the annular and concentric FS regimes. 

\begin{widetext}
Fermions have dispersion $\varepsilon_{j} = \vj \kpp + \frac{\kp^{2}}{2\mj}$ for patch $j$ at angle $\theta$ and couple via $\lambda_{j}$ to the Goldstone mode with the opposite sign $\text{sgn}(\lambda_{j})= -\text{sgn}(\lambda_{j'})$ between patches in a pair. Setting external momenta and frequencies to zero, the vertex correction gives:
\be
\bsplit
\delta u &= u \lambda_{j} \lambda_{j'} \int_{>} \frac{d\om d\pp d\ppp}{(2\pi)^{3}} G^{j}(\om,\bs{k})G^{j'}(\om,\bs{k}) D(\om,\bs{k})\\
&= \frac{iu}{N}  \lambda_{j} \lambda_{j'} \int_{>} \frac{d\om d\pp}{(2\pi)^{2}} \frac{\Theta(\om/\vj)-\Theta(\om/\vjp)}{\left[ \frac{\pp^{2}}{2}( \frac{\vj}{\mjp} - \frac{\vjp}{\mj}) - \frac{i\text{sgn}(\om)}{N} (\frac{\vj}{\zeta'} - \frac{\vjp}{\zeta}) |\om|^{\frac{2}{\zb}}  \right] \left[ \gamma \frac{|\om|}{|\pp|} + c^{\zb-1}|\pp|^{\zb-1} \right]} 
\end{split}
\ee
\end{widetext}

For the patches under consideration, the velocities $\vj, \vjp$ have opposite sign while $\mj,\mjp$ have the same sign. Defining $\alpha \equiv \frac{|\vj|}{\zeta'} + \frac{|\vjp|}{\zeta}$ and $\beta \equiv \frac{|\vj|}{|\mjp|} + \frac{|\vjp|}{|\mj|}$, one obtains:
\be
\bsplit
\frac{d(\delta u)}{dl} =& - u \frac{\lambda_{j} \lambda_{j'}}{\pi^{2} \gamma N}  \left[ \frac{|\vj|}{|\mjp|}+\frac{|\vjp|}{|\mj|} \right]^{-1} \times g(x,\zb) \\
g(x,\zb) \equiv & \int_{0}^{\infty} dt \frac{x t^{2/\zb}}{(x^{2} + t^{4/\zb})(1+t)}\\
x \equiv& \frac{N\beta \gamma^{2/\zb}}{2 \alpha c^{2(1-1/\zb)}} > 0
\end{split} 
\ee
Both $\zeta, \zeta'$ have a leading linear in $\epsilon$ behavior so $ x \sim \epsilon N$. Note that ${d(\delta u)}/{dl} > 0$ so correlations will be enhanced.

The Goldstone mode also mediates repulsive interactions for other patch pair types, e.g. (1,3), (2,4) for the annular FS or (1,2), (3,4) for the concentric FS. In principle we would be interested in these on grounds that they might be enhanced. However, here patch pairs have the same sign velocity, giving rise to $\ppp$ poles on the same side of the complex plane when external lines carry no frequency. Finite values in the external lines would contribute to a renormalization of $u(\bs{k},\omega)$ with finite arguments, which is less relevant than $u(0,0)$.

\subsubsection{$k^>_F \pm k^<_F$ Cooper channel susceptibility}
Similarly, we compute the vertex correction to the coupling $S_{\text{ext}} = \int d^{2}\bs{x} d\tau \left[ u_a \psi^{\dag}_{j} \psi^{\dag}_{j'} + h.c. \right]$ with $a=k^>_F \pm k^<_F$. This yields the susceptibility scaling form for superconducting order within patch pairs $(j,j') = (1,2)$ and $(3,4)$ for the annular FS and $(1,3)$ or $(2,4)$ for the concentric FS.  

\begin{widetext}
The coupling to the Goldstone mode has the same sign $\text{sgn}(\lambda_{j})= \text{sgn}(\lambda_{j'})$ for patches in a pair. Setting external momenta and frequencies to zero gives: 
\be
\bsplit
\delta u &= u \lambda_{j} \lambda_{j'} \int_{>} \frac{d\om d\pp d\ppp}{(2\pi)^{3}} G^{j} (\om,\bs{k})G^{j'}(-\om,-\bs{k}) D(\om,\bs{k})\\
&= +  \frac{iu}{N} \lambda_{j} \lambda_{j'} \int_{>} \frac{d\om d\pp}{(2\pi)^{2}} \frac{\Theta(\om/\vj)-\Theta(\om/\vjp)}{\left[ \frac{\pp^{2}}{2}( \frac{\vj}{\mjp} + \frac{\vjp}{\mj}) + \frac{i\text{sgn}(\om)}{N} |\om|^{\frac{2}{\zb}} (\frac{\vj}{\zeta'} - \frac{\vjp}{\zeta}) \right] \left[ \gamma \frac{|\om|}{|\pp|} + c^{\zb-1}|\pp|^{\zb-1} \right]} 
\end{split}
\ee
\end{widetext}
For the patches under consideration, the velocities $\vj, \vjp$ are opposite in sign while $\mj,\mjp$ may have opposite (annular FS) or the same sign (circular FS). Define $\alpha \equiv \frac{|\vj|}{\zeta'} + \frac{|\vjp|}{\zeta}$ and $\beta \equiv \left| \frac{|\vj|}{|\mjp|} \pm \frac{|\vjp|}{|\mj|} \right|$, where the $\pm$ is for the annular/concentric FS cases. The computation is otherwise the same as for the particle-hole channel and yields: 
\be
\bsplit
\frac{d(\delta u)}{dl} &= u \frac{\lambda_{j} \lambda_{j'}}{\pi^{2} \gamma N}  \left| \frac{|\vj|}{|\mjp|} \pm \frac{|\vjp|}{|\mj|}  \right|^{-1} \times g(x,\zb) 
\end{split}
\ee
with $x, g(x,\zb)$ defined as before. Note again that $d(\delta u)/dl >0$.

\subsubsection{Evaluation}
Within this RG scheme, we can take the limit $\epsilon \rightarrow 0, N \rightarrow \infty$ with $\epsilon N$ finite. For both the particle-hole and Cooper channels, we find a modified scaling dimension $u'_a = b^{\phi_{u_a}} u_a$ (including the bare value $\phi_{u_a} = 1$) to leading order in $\epsilon$:
\be
\bsplit
\phi_{u_a} = 1 + \epsilon \frac{ |\lambda_{j} \lambda_{j'}|}{ \pi^{2} \gamma \epsilon N} \left| \frac{|\vj|}{|\mjp|} \pm \frac{|\vjp|}{|\mj|}  \right|^{-1} g(\bar{x},\zb=2)\\
\end{split}
\ee
\be
\bsplit
g(x,&\zb=2) = \frac{\pi x^{2}}{2(1+x^{2})} \left[ 1- \frac{2}{\pi x} \log(x) \right]\\
\bar{x} \equiv& \lim_{\stackrel{ N \rightarrow \infty}{\epsilon \rightarrow 0}} x = (\epsilon N) \pi^{2} \gamma \left[ \frac{\lambda^{2}_{j'}|\vj|}{|\vjp|} + \frac{\lambda^{2}_{j}|\vjp|}{|\vj|} \right]^{-1} \left| \frac{|\vj|}{|\mjp|} \pm \frac{|\vjp|}{|\mj|} \right|
\end{split}
\ee

\begin{widetext}
Here, the lower sign is for the concentric FS Cooper channel; the upper sign applies otherwise. This gives a scaling form $\chi_a(\pp,\ppp,\omega) = |\omega|^{\delta_a} F_a(|\omega|/|\pp|^{\zb}, \ppp/\pp^{2})$ with exponent
\be
\bsplit
\delta_a = \frac{1}{2} + \frac{\epsilon}{4} \left[ 1- \frac{8 |\lambda_{j} \lambda_{j'}|}{\pi^{2} \gamma}  \left| \frac{|\vj|}{|\mjp|} \pm \frac{|\vjp|}{|\mj|}  \right|^{-1} \frac{g(\bar{x},\zb=2)}{\epsilon N} \right] + \mathcal{O}(\epsilon^{2})
\end{split}
\ee
(Note the channel $a$ is dependent on patch pair $(j,j')$ being considered).
\end{widetext}

\begin{center}
\begin{figure}[t]
\centering
\includegraphics[width = 2.7in]{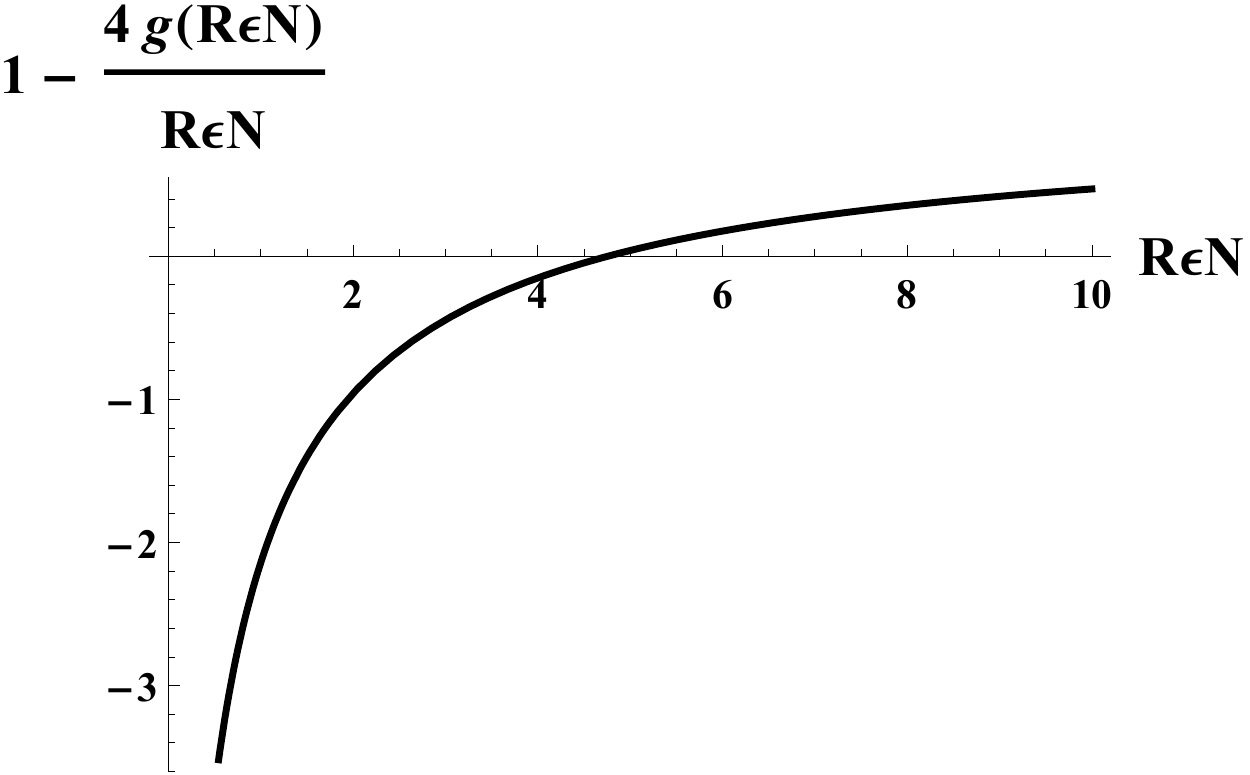} 
\caption{Plot of $\left[ 1 - 4 \cdot \frac{ g(R_a \epsilon N, \zb=2)}{R_a \epsilon N} \right]$ (with channel dependence $a$) as a function of $R_a \epsilon N$ which leads to a singular susceptibility for sufficiently negative values. The zero crossing is at $R_a \epsilon N \approx 4.8$ and at large $R_a \epsilon N$ the asymptote is 1.}
\label{fig:GeneralNeps}
\end{figure}
\end{center}

As an example, we simplify to the most important case $\theta = 0$, $|\vj| = |\vjp| \equiv v, |\lambda_{j}| = |\lambda_{j'}| \equiv \lambda$ and masses $|\mone|=|m_{4}|,|\mtwo|=|m_{3}|$. These patches couple most strongly to the Goldstone mode and hence will receive the strongest singular enhancemement of susceptibilities. In this case, $\gamma = \lambda^{2} (|\mone|+|\mtwo|)/(2\pi v)$. It is convenient to define a dimensionless ratio $R_a$ for each channel $a$:
\be
\bsplit
R_a \equiv \frac{\pi}{4} \frac{(|\mone|+|\mtwo|)\left| |\mj| \pm |\mjp| \right|}{|\mj||\mjp|}\\
\end{split}
\ee
where $\mj,\mjp$ are chosen based on the channel of interest (lower sign is for concentric FS Cooper channel, upper sign otherwise).  Expressions for $R_a$ for the various channels are given in Table 
\ref{table:ratios} as a function of the mass ratio $r_{M}= |\mtwo|/|\mone| \in (0,1)$.  Physically, $R_a$ can be interpreted as the ratio of the effective density of states (DOS) for a given channel $\sim |\frac{1}{|\mj|} \pm \frac{1}{|\mjp|}|^{-1} $ to the total DOS $\sim (|\mone| + |\mtwo|)$ that enters the Landau damping coefficient. Qualitatively, Fig. \ref{fig:GeneralNeps} shows that larger $R_a$ values lead to weaker power law susceptibilities.  

For general $N$ and $\epsilon$, the value of the susceptibility exponent $\delta_a$ is:
\be
\bsplit
\delta_a = \frac{1}{2} + \frac{\epsilon}{4} \left[ 1 - 4 \cdot \frac{ g(R_a \epsilon N, \zb=2)}{R_a \epsilon N} \right]
\label{eq:GeneralCase}
\end{split}
\ee
A plot of the coefficient of $\epsilon/4$ is shown in Fig. \ref{fig:GeneralNeps}. The physically interesting limit $N, \epsilon=1$ lies beyond the controlled regime explored above; we now extrapolate the above results to this case. We still consider the most important case $\theta=0$. With $R_a$ defined as above, the power law in the scaling form is
\be
\bsplit
\delta_a &= \frac{2}{3}\left[1 - \frac{g(\sqrt{3}R_a/\pi,\zb=3)}{R_a} \right]
\label{eq:SpecificCase}
\end{split}
\ee
We plot $\delta_a$ as a function of $x \equiv \kf/\ko$ for all 4 channels in Fig. \ref{fig:SpecificNeps}, where $r_{M} = |x - 1|/(x+1)$. At the angle $\theta=0$, $x = \sqrt{2\mu}/(\sqrt{\mo} \alpha_R)$ and therefore increases monotonically with $\mu$. $x = 1$ denotes the Dirac point. 

\begin{center}
\begin{table}[t]
\renewcommand{\arraystretch}{1.4}
\begin{tabular}{|c|c|c|c|}
\hline
SDW (1,4) & SDW (2,3) & CP annular FS & CP concentric FS\\
\hline
$\frac{\pi}{2}(1+r_{M})$ & $\frac{\pi}{2}(1+\frac{1}{r_{M}})$ & $\frac{\pi}{4}\frac{(1+r_{M})^2}{r_M}$ & $\frac{\pi}{4}\frac{(1-r_{M}^2)}{r_M}$\\
\hline
\end{tabular}
\caption{Value of dimensionless mass ratio $R_a$ as a function of $r_{M} = |\mtwo|/|\mone| \in (0,1)$.}
\label{table:ratios}
\end{table}
\end{center}

Only the susceptibility in the Cooper channel with concentric FS has a power law divergence for sufficiently high doping in the Rashba liquid (the zero crossing is at $x \approx 4.6$). We note the Cooper pairing here is within patch pairs $(j,j') = (1,3)$ and $(2,4)$. While the difference in the radii of the two concentric Fermi surfaces stays finite as $\mu$ increases, the masses of the two patches approach the same value, mimicking the usual time-reversal symmetric BCS pairing arising from nested scattering.

All the curves except for the particle-hole channel $(1,4)$ peak at $x = 1$ when $\mtwo = 0$. One can see that for these other channels, $R_a$ depends on $1/r_{M}$ and $R_a \rightarrow \infty$ as $|\mtwo| \rightarrow 0$. This is because the ``effective" DOS, which is some reduced combination of the masses of two patches, is vanishing. For the particle-hole $(1,4)$ channel, $R_a$ instead takes it minimal value at $x = 1$.

\subsubsection{Special Case: TI Surface}
The TI surface can be considered as a special case of the above results, where only a single SDW channel may occur. The TI Landau damping coefficient reads $\gamma = \lambda^{2} \mu / (2\pi \vd^{3})$ and for $\epsilon \rightarrow 0, N \rightarrow \infty, \epsilon N$ finite, Eq. \ref{eq:GeneralCase} holds with $R_{2k_F} = \pi/2$, choosing the upper sign. For the case $N, \epsilon = 1$, Eq.~\ref{eq:SpecificCase} applies with $R_{2k_F}=\pi/2$ giving $\gamma \approx 0.2$. That is, within the $(N, \epsilon)$ approximation the TI surface state is stable against SDW formation.

\bibliographystyle{apsrev}  
\bibliography{Rashba_refs,bib}

\begin{thebibliography}{41}
\bibitem[{Dia()}]{DiamagneticTerm}
\bibinfo{note}{Quadratic contributions\cite{MaxIsingI,NEpsilon} (analogous to
  diamagnetic terms for the problem of a Fermi surface coupled to a gauge
  field) are automatically incorporated by a proper choice of order of
  integrations}.
\bibitem[{RPA()}]{RPA}
\bibinfo{note}{Typically, RPA treatments are justified by a large $N$
  expansion. Here, the large $N$ expansion is known to break down.\cite{SSLee2}
  The use of RPA is formally justified by a more complicated double
  expansion\cite{NEpsilon} and also by uncontrolled higher loop
  calculations\cite{MaxIsingI} which show that $z_b=3$ RPA scaling is valid
  through three loop order.}
\end{thebibliography}

\end{document}